\documentclass[aps,prx,reprint,superscriptaddress,longbibliography]{revtex4-2}
\usepackage{times}
\usepackage{amsmath,amssymb,amsthm,amsfonts,bm,braket,dcolumn,graphicx,mathrsfs,multirow,physics,xcolor,txfonts}
\usepackage{soul}
\usepackage[pdftex,colorlinks,linkcolor={red!75!black},citecolor={red!75!black},urlcolor={blue!75!black}]{hyperref}
\usepackage{comment}
\usepackage{xcolor}
\usepackage{xcolor}
\usepackage[normalem]{ulem}

\newcommand{\clr}{\color{red!75!black}}

\newcommand{\Z}{\mathbb{Z}}

\begin{document}

\title{Algebraic Relaxation of Strong-to-Weak Symmetry-Breaking Order via Charge-Sector Broadening}

\author{Zi-Xi Fang}
\affiliation{School of Physics and Zhejiang Key Laboratory of Micro-nano Quantum Chips and Quantum Control, Zhejiang University, Hangzhou 310058, China}
\affiliation{Beijing National Laboratory for Condensed Matter Physics, and Institute of Physics, Chinese Academy of Sciences, Beijing 100190, China}

\author{Kai Zhang}
\email{phykai@zju.edu.cn}
\affiliation{School of Physics and Zhejiang Key Laboratory of Micro-nano Quantum Chips and Quantum Control, Zhejiang University, Hangzhou 310058, China}
\date{\today}

\begin{abstract}
    Strong-to-weak spontaneous symmetry breaking (SW-SSB) is a distinctive phenomenon of mixed quantum states, characterized by nonlinear long-range order associated with strong symmetry. In realistic open quantum systems, however, strong symmetry is macroscopically fragile and can be explicitly reduced to weak symmetry through local charge exchange with the environment. 
    In this paper, we investigate how the SW-SSB order decays under such strong-symmetry-breaking Liouvillian dynamics. For discrete $\mathbb{Z}_2$ symmetry, a local symmetry-breaking perturbation drives an exponential decay of the R\'enyi-1 long-range correlator, revealing the fragility of discrete SW-SSB order. 
    By contrast, for continuous $U(1)$ symmetry, local charge exchange broadens the distribution over symmetry-charge sectors and leads to parametrically slower algebraic relaxation. We establish a rigorous algebraic lower bound on the connected R\'enyi-1 long-range correlation, demonstrating the robustness of this slow relaxation. In the thermodynamic limit, we further uncover a two-stage dynamical scaling: $\tilde R_1\sim t^{-1}$ at early times and $\tilde R_1\sim t^{-1/z}$ at long times, with $z$ set by the underlying hydrodynamic dynamics. As a numerical verification, we consider fermionic chains with $z=2$ and $z=4$, which are excellent agreement with the theoretical predictions. 
\end{abstract}

\maketitle

\emph{{\clr Introduction}.---}
For pure quantum states, symmetry acts directly on state vectors and provides a fundamental organizing principle for quantum many-body phases. For mixed quantum states, however, symmetry can be realized in two distinct forms~\cite{Prosen2012IOP,JiangL2014PRA,Coser2019Quantum,Gorshkov2020PRL,Schuch2022Quantum,XuCK2023PRXQ,WangC2025PRX,Lessa2025PRX,BiZhen2025PRX,XQSun2025PRL,IKuo2024PRB,XueJia2025PRXQuantum}. A density matrix possesses a strong symmetry when every state in ensemble decomposition carries the same symmetry charge, whereas a weak symmetry requires only invariance of the density matrix as an ensemble. 
This distinction has led to the recently developed concept of strong-to-weak spontaneous symmetry breaking (SW-SSB), where conventional linear correlators remain short-ranged while nonlinear fidelity or R\'enyi correlators develop long-range order~\cite{Lessa2025PRXQ,Weinstein2025PRL,ZhangPF2025CP,FengL2025PRL,Ippoliti2025PRL,YouYZ2024PRB,WangZhong2025PRB,ShuoYang2025PRB,Ziereis2025arXiv,WangChong2026Review,JHZhang2024,IBloch2026,YZYou2026PRXQ,MFisher2026arXiv,Chang2026arXiv,Carolyn2026arXiv,Haga2026arXiv,PFZhang2025PRB}. 
These nonlinear correlators also have a direct information-theoretic interpretation~\cite{HsiehT2023PRXQ,Sang2024PRX,Hsieh2025PRL,Lessa2025PRXQ,Verresen2023PRL,MatthewF2023PRB,JiangL2024PRL,ZhangYF2024PRA,ZhangYF2025PRL,LeeJY2025PRL,Timothy2025,JongYLee2026arXiv,TimothyHH2026arXiv,Wang2026LocalSWSSB,LiuRZ2026arXiv,DongELiu2026PRB}: the fidelity/R\'enyi-1 correlator quantifies the distinguishability between a mixed state and one obtained by moving symmetry charge over a long distance~\cite{Lessa2025PRXQ,Weinstein2025PRL,ZhangPF2025CP}. 
Their long-range order signals that globally sharp symmetry information is stored nonlocally and becomes inaccessible to local recovery, as reflected in long-range conditional mutual information and related notions of charge scrambling and learnability~\cite{Omar2015CMP,Hsieh2025PRL,Vasseur2022PRX,JongYLee2026arXiv,TimothyHH2026arXiv,Knap2025arXiv}.
This phenomenon has recently gained experimental relevance through the observation of long-range R\'enyi correlations in a dephased Fermi gas~\cite{IBloch2026}. 

The formulation of SW-SSB, however, requires an exact strong symmetry, such that the density matrix has a definite global symmetry charge. In a realistic open system, however, even weak local coupling to an environment can permit symmetry charge exchange~\cite{Lessa2025PRXQ,XQSun2025PRL,WangChong2026Review}, thereby reducing a strong symmetry to a weak counterpart, analogous to the distinction between canonical and grand-canonical ensembles. 
Once symmetry-charge exchange breaks the exact strong symmetry, SW-SSB order is no longer formally defined. Nevertheless, this does not imply that the nonlinear long-range correlations associated with SW-SSB order must disappear immediately. 

This distinction exposes a basic stability problem: what happens to SW-SSB order after that strong symmetry is explicitly broken? 
Previous work has shown that SW-SSB states exhibit a divergent fidelity susceptibility to spatially extensive strong-symmetry-breaking quantum channels~\cite{JHZhang2024,WangChong2026Review}. Such a thermodynamic susceptibility, however, does not address the qualitatively different situation in which strong symmetry is violated only within a finite spatial region [Fig.~\ref{fig:1}(a)]. 
A local coupling directly exchanges only an $O(1)$ amount of symmetry charge with the system, while the order it perturbs is encoded nonlocally throughout the many-body state. This setting is also accessible to experiment~\cite{IBloch2026}, where nonlinear correlations can be directly measured even though the exact strong is broken. 
These considerations raise a natural question: can nonlinear long-range order survive an explicit local strong-symmetry breaking? If so, how does it decay, and what controls this decay? 

This work addresses this stability problem by investigating the relaxation of nonlinear long-range order under strong-symmetry-breaking Liouvillian dynamics. We uncover sharply distinct relaxation mechanisms for discrete $\mathbb{Z}_2$ and continuous $U(1)$ symmetries. 
For $\mathbb{Z}_2$, a local symmetry-breaking channel mixes the two global symmetry-charge sectors at a finite rate, leading directly to an exponential decay of the long-range R\'enyi-1 order~[Fig.~\ref{fig:1}(b)(c)]. 
For $U(1)$, by contrast, the information is encoded in the sharpness of an entire distribution over total symmetry charge sectors. Local symmetry-breaking channels gradually broaden this distribution through a diffusion-like process in symmetry-charge space, producing a parametrically slower algebraic decay of the connected nonlinear R\'enyi-1 order~[Fig.~\ref{fig:2}]. 

For the U(1) case, we derive a rigorous algebraic slower bound on the connected R\'enyi-1 correlation and identify its universal two-stage dynamical relaxation scalings. 
In the thermodynamic limit, the R\'enyi-1 long-range correlation initially decays as $t^{-1}$, reflecting the diffusive broadening of the global U(1) symmetry-charge distribution induced by symmetric simple exclusion process~\cite{Mallick2015,Bernard2018,Wellnitz2022PRL}, and universally cross over to $t^{-1/z}$ at long times, where $z$ is the dynamical exponent governing the underlying $U(1)$ hydrodynamics~\cite{Knap2020PRL,JoneYeon2023PRL,Moudgalya2024PRXQ,Lucas2025PRB,WangZhong2025PRB}. 
Our results show that explicit local breaking of strong symmetry need not immediately erase the information-theoretic order associated with SW-SSB, and provide experimentally accessible dynamical scaling of its robustness beyond the idealized limit of exact strong symmetry. 

\begin{figure}[t]
    \begin{centering}
    \includegraphics[width=\linewidth]{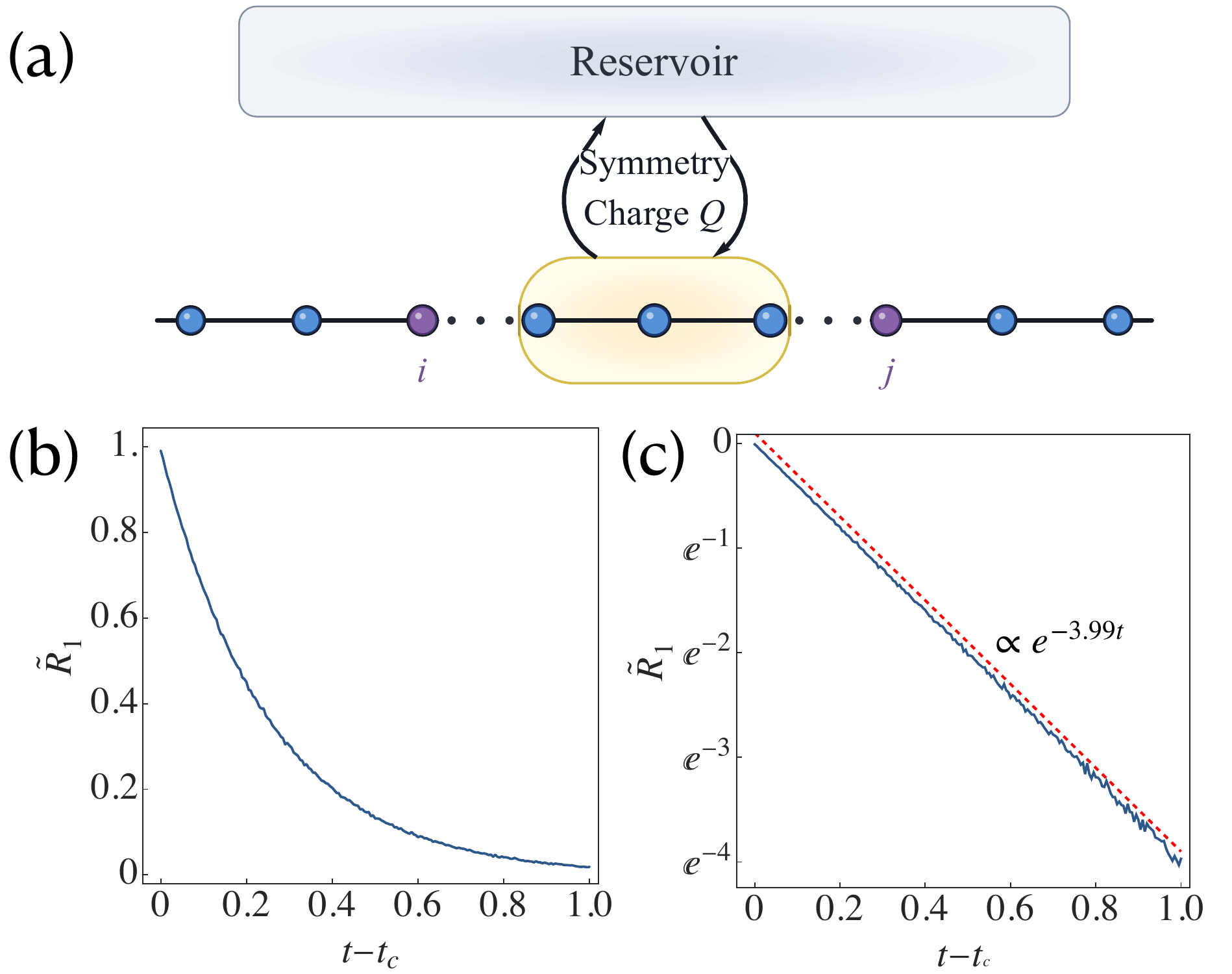}
    \par\end{centering}
    \protect\caption{\label{fig:1} 
    (a) Illustration of local strong-symmetry breaking on a finite lattice region. (b) Exponential decay of the long-range connected R\'enyi-1 correlator under the local symmetry-breaking perturbation in Eq.~\eqref{EQ_Z2SymmBreaking}, with a semilogarithmic plot shown in (c). The unperturbed strongly $\mathbb Z_2$-symmetric Liouvillian is $\mathcal L_0[\rho]=\gamma_x\sum_{i=1}^{L}\mathcal D[X_i]\rho+\gamma_z\sum_{i=1}^{L-1}\mathcal D[Z_iZ_{i+1}]\rho+\gamma_4\sum_{i=1}^{L-3}\mathcal D[Z_iZ_{i+1}Z_{i+2}Z_{i+3}]\rho$, where $\mathcal D[O]\rho=O \rho O^\dagger-\frac12\{O^\dagger O,\rho\}$.
    Parameters are $(\kappa,\gamma_x,\gamma_z,\gamma_4)=(2,1,1,0.3)$. }
\end{figure}

\emph{{\clr Connected R\'enyi-k correlation without exact strong symmetry}.---}
Consider a strongly symmetric Liouvillian $\mathcal{L}_0$ subject to a generic local symmetry-breaking perturbation $\mathcal{V}_R$, supported on a finite spatial region $R$ in the thermodynamic limit [Fig.~\ref{fig:1}(a)]. 
The perturbation explicitly reduces the strong symmetry to its weak counterpart. Physically, $\mathcal{V}_R$ describes coupling a single site or a finite local region to an external reservoir, thereby allowing symmetry charge to be exchanged with the environment. 
The perturbation is switched on at time $t_c$, such that the dynamics is governed by
\begin{equation}\label{EQ_GeneralLiou}
    \frac{d\rho(t)}{d t} = \mathcal{L}_0[\rho(t)] + \Theta(t-t_c)\, \mathcal{V}_R[\rho(t)],
\end{equation}
where $\Theta(t)$ denotes the Heaviside step function.
Note that the evolution time can equivalently be interpreted as the depth of the corresponding quantum circuit~\cite{SupMat}. 

To diagnose the nonlinear long-range order that survives the explicit breaking of strong symmetry, we introduce the connected R\'enyi-$k$ correlator
\begin{equation}\label{EQ_ConnectRenyik}
    \begin{split}
    \tilde{R}_k(\rho,O_i O^\dagger_j) = R_k(\rho,O_i O^\dagger_j) - R_k(\rho,O_i) R_k(\rho,O_j)
    \end{split}
\end{equation}
where $R_k(\rho,O) \equiv \Tr[\rho^{k/2} O \rho^{k/2} O^\dagger] / \Tr[\rho^k]$, and $O_{i,j}$ are local charged operators carrying nonzero symmetry charge, namely $U_g O_i U_g^\dagger=\chi_q(g)O_i$ with $\chi_q(g)\neq 1$. 
When $\rho$ has exact strong symmetry, $R_k(\rho,O_i)=0$ for any positive integer $k$~\cite{footnote1}, and Eq.~\eqref{EQ_ConnectRenyik} reduces to the usual nonlinear diagnostic of SW-SSB order~\cite{IBloch2026}. 
Once the strong symmetry is explicitly broken, the one-point correlation $R_k(\rho, O_i)$ is generally nonzero. 
The connected correlator subtracts this local contribution and therefore presents the residual long-range correlations inherited from the SW-SSB state. 
We focus on the R\'enyi-1 correlator throughout the main text, while further discussions of the connected R\'enyi-k correlator and conditional mutual information are presented in Sup.~Mat.~C and D, respectively~\cite{SupMat}. 

\emph{{\clr Exponential decay in $\mathbb{Z}_2$ strong-symmetry-breaking dynamics}.---}
We first consider Lindblad dynamics with a strong $\mathbb{Z}_2$ symmetry. The unperturbed evolution is generated by nearest-neighbor dephasing
\begin{equation}\label{EQ_ZZDephasing}
    \mathcal{L}_0[\rho]=\gamma \sum_i \left(Z_i Z_{i+1}\rho Z_{i+1}Z_i-\rho\right),
\end{equation}
where $Z_i$ is the Pauli-$z$ operator acting on site $i$. 
This Liouvillian is strongly symmetric under the global spin-flip operator $X=\prod_i X_i$, since $[Z_iZ_{i+1},X]=0$. 
Consequently, states in a definite parity sector remain in that sector under $\mathcal{L}_0$ with steady states $\rho_\pm = (\mathbb{I} \pm X)/2^{L}$, where $L$ is the lattice size. 

To directly probe the fate of the nonlinear SW-SSB order under explicit $\Z_2$ symmetry breaking, we choose the initial state $\rho_0=\rho_+$ that exhibits SW-SSB order characterized by a long-range R\'enyi-1 correlation $R_1(\rho_+,Z_iZ_j)=1$. 
Evolve it for $t>t_c=0$ under $\mathcal L_0 + \mathcal V_x$, where 
\begin{equation}\label{EQ_Z2SymmBreaking}
    \mathcal{V}_x[\rho] = \frac{\kappa}{2}(Z_x \rho Z_x - \rho),
\end{equation}
where $\kappa$ is the strength of symmetry-breaking channel. 
Since $X Z_x X = -Z_x$, the perturbation breaks the strong $\Z_2$ symmetry down to its weak counterpart and connects the two symmetry charge sectors. 
Moreover, $\mathcal{L}_0[\rho_\pm]=0$ and $Z_x \rho_{\pm} Z_x = \rho_{\mp}$, so the dynamics remains within the manifold spanned by $\rho_{\pm}$, namely, $\rho(t) = w_+(t) \rho_+ + w_-(t)\rho_-$ with $w_+(t) + w_-(t)=1$. 
Define the charge-sector imbalance $a(t) = w_+(t) - w_-(t)$, and it evolves under the symmetry-breaking dynamics Eq.~\eqref{EQ_GeneralLiou} as $\dot{a}(t)=-\kappa a(t)$. 
The initial state has definite symmetry charge, thus $a(t=0)=1$. 
It follows that $a(t)=e^{-\kappa t}$ and therefore, $w_{\pm}(t) = (1\pm e^{-\kappa t})/2$. The state thus evolves as $\rho(t) = \frac{1+e^{-\kappa t}}{2} \rho_+ + \frac{1-e^{-\kappa t}}{2} \rho_-$. 
For this $\Z_2$ example, the connected R\'enyi-$1$ correlation is calculated as 
\begin{equation}
    \lim_{|i-j|\to \infty} \tilde{R}_1(\rho(t),Z_iZ_j) = e^{-2 \kappa t},
\end{equation}
which shows that the nonlinear long-range order decays exponentially under local strong-symmetry breaking, thus indicating the fragility of SW-SSB order in discrete symmetry class. 
The same exponential fragility persists in the generic strongly \(\mathbb Z_2\)-symmetric model shown in Figs.~\ref{fig:1}(b)(c), demonstrating that it does not rely on the solvable structure of Eq.~\eqref{EQ_ZZDephasing}. 

\emph{{\clr Algebraic relaxation in $U(1)$ strong-symmetry-breaking dynamics}.---}
We now consider a generic fermionic Lindblad dynamics with strong U(1) symmetry, generated by an unperturbed Liouvillian $\mathcal L_0$ that conserves the total particle number $Q=\sum_i c_i^\dagger c_i$. 
We locally break the strong symmetry by coupling a single site $x$ to a particle reservoir,
\begin{equation}\label{EQ_U1SymmBreaking}
	\mathcal V_x[\rho(t)] = \kappa_- \mathcal{D}[c_x] \rho(t) + \kappa_+ \mathcal{D}[c_x^\dagger] \rho(t),
\end{equation}
where $\kappa_-$ and $\kappa_+$ denote the particle-loss and particle-gain rates, respectively.
Although we consider the symmetry-breaking channel $V_x$ acting on a single site for demonstration, the underlying mechanism can be straightforwardly generalized to any finite region in the thermodynamic limit.
While $\mathcal L_0$ preserves each fixed-charge sector, $\mathcal V_x$ exchanges particles locally with the reservoir and therefore transfers charge-sector weight between neighboring sectors $Q \leftrightarrow Q \pm 1$. 
The strong $U(1)$ symmetry is thus explicitly reduced to its weak counterpart. 

For the analytical discussion, it is convenient to choose the initial state as a fixed point of $\mathcal L_0$. 
We denote the maximally mixed state (MMS)~\cite{Ziereis2025arXiv,JoneYeon2023PRL} in the charge-$Q$ sector by $\bar P_Q=P_Q/D_Q$, where $D_Q=\mathrm{Tr}[P_Q]$ and $P_Q$ projects onto that sector. 
The state $\bar P_Q$ has exact strong $U(1)$ symmetry and exhibits long-range SW-SSB order, characterized by a nonvanishing long-range R\'enyi-1 correlator. 
We therefore take $\rho(0)=\bar P_Q$, which is prepared to exhibit the effect of the subsequent strong-symmetry-breaking dynamics on the preexisting nonlinear long-range order. 

Because the full evolution preserves weak $U(1)$ symmetry, an initially charge-diagonal state remains block diagonal in charge space and can be written generally as
\begin{equation}
    \rho(t)=\sum_Qw_Q(t)\rho_Q(t),
\end{equation}
where $w_Q(t)=\mathrm{Tr}[P_Q\rho(t)]$ is the charge-sector weight and $\rho_Q(t)=P_Q\rho(t)P_Q/w_Q(t)$ is the normalized state within sector $Q$. 
Starting from the sharp distribution $w_Q(0)=\delta_{Q,Q_0}$ in the strong symmetry limit, the local particle exchange generated by $\mathcal{V}_x$ broadens the distribution over the ladder of $U(1)$ charge sectors~[Fig.~\ref{fig:2}(a)].
In contrast to the $\Z_2$ case, where the sector dynamics is fully characterized by a single imbalance $w_+(t)-w_-(t)$, the $U(1)$ dynamics involves the entire distribution $w_Q(t)$, thus the charge-sector variance $\sigma_Q^2(t)$ provides a natural measure of charge-space broadening. 

\begin{figure}[t]
    \begin{centering}
    \includegraphics[width=1\linewidth]{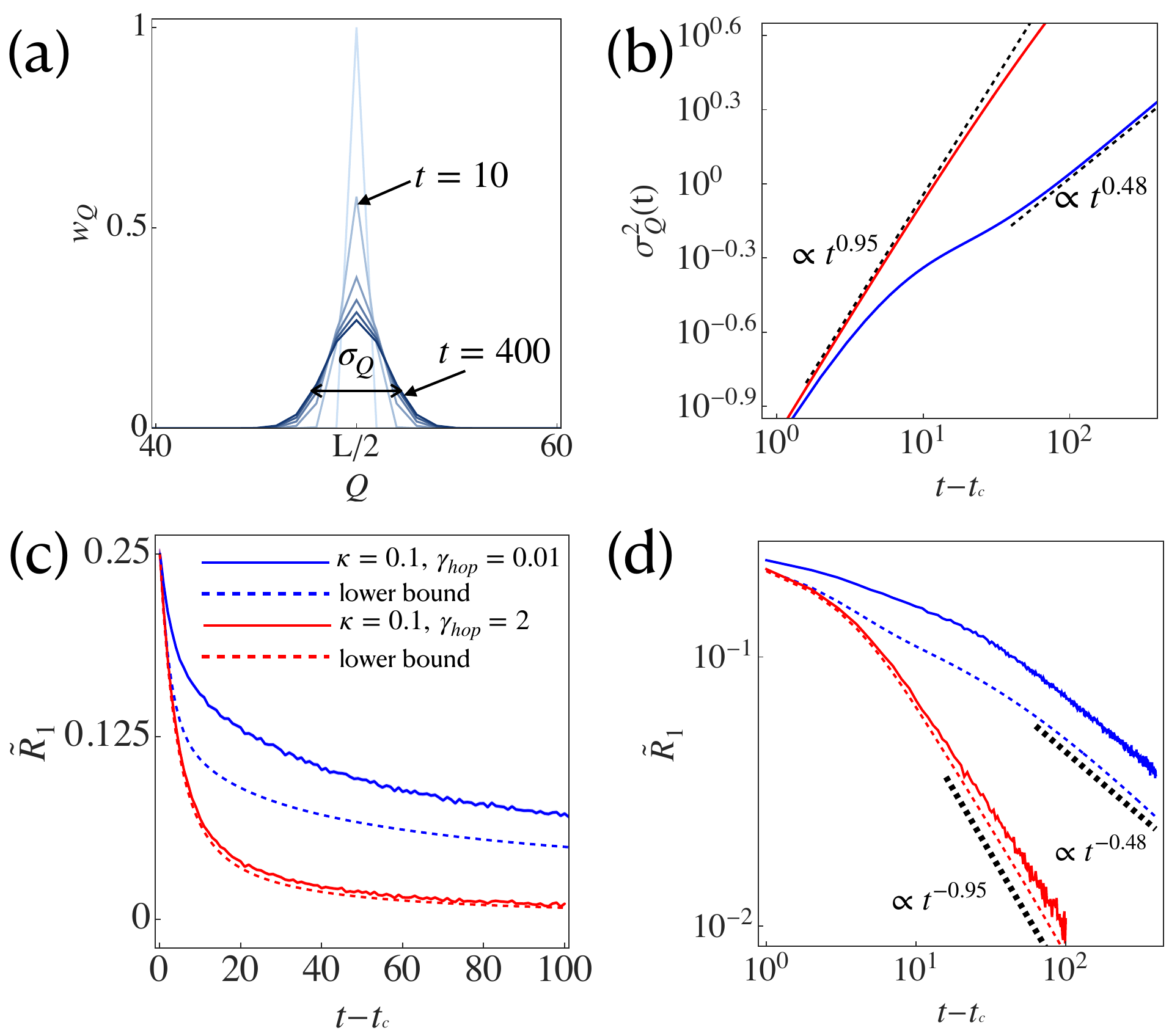}
    \par\end{centering}
    \protect\caption{\label{fig:2}
    \textbf{Charge-space broadening and R\'enyi-1 algebraic relaxation under local $U(1)$ charge exchange.} Numerical results for a spinless-fermion chain in Eq.~\eqref{EQ_simulation_model} with $L=100$ and $z=2$, initialized in $\bar P_{L/2}$. A balanced leakage channel with $\kappa_+=\kappa_-=\kappa=0.1$ acts at $x=L/2$, with $\gamma_\phi=5$. Blue and red denote $\gamma_{\mathrm{hop}}=0.01$ and $2$, respectively. 
    (a) Broadening of the charge-sector distribution $w_Q(t)$ for $\gamma_{\mathrm{hop}}=0.01$, with width $\sigma_Q$. 
    (b) Charge variance on logarithmic scales; black dashed lines indicate $\sigma_Q^2\propto(t-t_c)^{0.48}$ (blue) and $(t-t_c)^{0.95}$ (red). 
    (c) Connected R\'enyi-1 correlator $\widetilde R_1(\rho,c_i^\dagger c_j)$ for $i=L/4$ and $j=3L/4$. Solid curves are numerical results, while colored dashed curves are the corresponding auxiliary-state predictions computed from the same $w_Q(t)$. 
    (d) Log-log plot of the same correlators; black dashed lines indicate the corresponding algebraic behaviors $(t-t_c)^{-0.48}$ and $(t-t_c)^{-0.95}$. The solid curves are estimated using $4\times10^5$ (blue) and $8\times10^5$ (red) Monte Carlo samples.}
\end{figure}

The relaxation of the connected R\'enyi-1 correlator is caused by the the loss of charge-sector sharpness. 
It's controlled not only by the distribution of the charge-sector weights $w_Q(t)$, but also by the intra-sector states $\rho_Q(t)$. 
The strong-symmetry-breaking channel $\mathcal V_x$ generally drives $\rho_Q(t)$ away from the MMS $\bar P_Q$. 
We therefore write $\rho_Q(t)=\bar P_Q+\delta_Q(t)$. The idealized limit $\delta_Q(t)=0$ corresponds to instantaneous intra-sector equilibration, in which the state relaxes back to $\bar P_Q$ immediately after each charge-leakage event. 

To capture the decay of connected R\'enyi-1 correlation, we introduce an auxiliary mixed-state
\begin{equation}\label{EQ:AuxMixedState}
    \rho_{\text{aux}}(t) \equiv \sum_Q w_Q(t) \bar P_Q,
\end{equation}
which is constructed by retaining the exact charge-sector weights $w_Q(t)=\Tr[\rho(t)P_Q]$ while replacing each normalized sector state $\rho_Q(t)$ with the corresponding charge-sector MMS $\bar P_Q$. 
We prove that the auxiliary state provides a lower bound on the long-range connected R\'enyi-1 correlator (see proof in Sup.~Mat.~C~\cite{SupMat}):
\begin{equation}\label{EQ_R1R1MMS}
	\lim_{|i-j|\to \infty} \widetilde R_1(\rho(t),c_i^\dagger c_j) \gtrsim \overline{\mathcal{F}^2}(t) \,\widetilde R_1(\rho_{\mathrm{aux}}(t),c_i^\dagger c_j),
\end{equation}
where $\overline{\mathcal{F}^2}(t) = \sum_Q w_Q(t) \mathcal{F}^2(\rho_Q(t),\bar P_Q)$ is the sector-averaged squared fidelity, with $\mathcal{F}(\rho,\sigma)=\mathrm{Tr}\sqrt{\sqrt{\rho}\sigma\sqrt{\rho}}$. 
This factor quantifies the average intra-sector similarity between $\rho_Q(t)$ and $\bar P_Q$ and satisfies $0<\overline{\mathcal F^2}(t)\leq 1$. 
Provided that $\overline{\mathcal F^2}(t)$ is lower bounded by an $O(1)$ constant $\overline{\mathcal{F}^2}(t)\geq A>0$, Eq.~\eqref{EQ_R1R1MMS} implies $\lim_{|i-j|\to \infty} \widetilde R_1(\rho(t),c_i^\dagger c_j) \geq A \,\widetilde R_1(\rho_{\mathrm{aux}}(t),c_i^\dagger c_j)$. 
The auxiliary-state R\'enyi-1 correlation therefore provides, up to a finite prefactor, an asymptotic lower bound on the connected R\'enyi-1 correlation. 

Since each charge-sector state is replaced with $\rho_Q=\bar P_Q$ in the auxiliary state, as given in Eq.~\eqref{EQ:AuxMixedState}, it evolves solely through the charge-sector weights $w_Q(t)$. 
The auxiliary-state R\'enyi-1 correlation can therefore be evaluated analytically as (see derivations in Sup.~Mat.~C~\cite{SupMat}):
\begin{equation}\label{EQ_R1mms}
	\lim_{|i-j|\to \infty}\widetilde R_1(\rho_{\mathrm{aux}}(t),c_i^\dagger c_j)\simeq \bar \nu(t)[1-\bar \nu(t)]\left[1-S_w^2(t)\right],
\end{equation}
where $S_w(t) = \sum_Q\sqrt{w_Q(t)w_{Q+1}(t)}$ is the Bhattacharyya overlap between the charge-sector distribution and its one-sector shift. 
$\bar\nu(t)=L^{-1}\sum_QQw_Q(t)$ denotes the mean filling, which replaces the fixed filling once the strong $U(1)$ symmetry is explicitly broken and the state spreads over multiple charge sectors. 
In the strongly symmetric limit, $w_Q(t=0)=\delta_{Q,Q_0}$ and hence $S_w(t=0)=0$. The R\'enyi-1 correlation therefore reduces to $\nu(1-\nu)$, reproducing the standard result for $U(1)$ SW-SSB~\cite{IBloch2026}. 

To determine the decay of auxiliary-state R\'enyi-1 correlation, we need to investigate how the charge-sector distribution $w_Q(t)$ evolves with time, as given by Eq.~\eqref{EQ_R1mms}. 
The local loss and gain processes generated by $\mathcal V_x$, connecting sector $Q$ to its neighboring sectors $Q-1$ and $Q+1$. 
Projecting the Lindblad equation onto sector $Q$, we obtain the evolution equation for the charge-sector weights (see derivations in Sup.~Mat.~A~\cite{SupMat})
\begin{equation}\label{EQ_U1ChargeEvolution}
	\frac{dw_Q(t)}{dt}=a_{Q+1}w_{Q+1}(t)+b_{Q-1}w_{Q-1}(t)-\left[a_Q+b_Q\right]w_Q(t),
\end{equation}
where $a_Q(t)=\kappa_-\mathrm{Tr}[n_x\rho_Q(t)]$ is the transition rate from $Q$ to $Q-1$, while $b_Q(t)=\kappa_+\left[1-\mathrm{Tr}[n_x\rho_Q(t)]\right]$ is the transition rate from $Q$ to $Q+1$. 
Although $\mathcal L_0$ does not directly transfer state between charge sectors, it governs the intra-sector states $\rho_Q(t)$ and thereby determines the time-dependent transition rates $a_Q(t)$ and $b_Q(t)$. These rates encode temporal correlations between successive charge-exchange events and control the evolution of the sector weights $w_Q(t)$. 
For the broad and smooth charge distribution generated by Eq.~\eqref{EQ_U1ChargeEvolution}, we approximate $w_Q(t)$ by a Gaussian distribution with variance $\sigma_Q^2(t)$. 
The charge-sector Bhattacharyya overlap in Eq.~\eqref{EQ_R1mms} then becomes
\begin{equation}\label{EQ_U1ChargeOverlap}
	S_w^2(t)\simeq1-\frac{1}{4\sigma_Q^2(t)},
\end{equation}
for $\sigma_Q^2(t)=\langle Q^2\rangle-\langle Q\rangle^2 \gg1$. 

We next determine the growth of the charge variance $\sigma_Q^2(t)$, which characterizes the broadening dynamics of the charge-sector distribution. 
Define $\kappa\equiv(\kappa_++\kappa_-)/2$, one can solve the dynamical equation in symmetry-charge space and obtain (see Sup.~Mat.~B~\cite{SupMat})
\begin{equation}\label{EQ_U1ChargeVarianceExact}
	\frac{d\sigma_Q^2(t)}{dt} = 4\kappa[\chi-C_x(t)]+(\kappa_--\kappa_+)\langle \delta \hat n_x(t)\rangle,
\end{equation}
where $C_x(t)=\mathrm{Tr}[\rho(t) \hat Q \hat n_x]-\mathrm{Tr}[\rho(t) \hat Q]\mathrm{Tr}[\rho(t) \hat n_x]$ is the covariance between the total charge and the local occupation at the symmetry-breaking site $x$, and $\chi=\nu_s(1-\nu_s)$, with $\nu_s=\kappa_+/(2\kappa)$ denoting the stationary filling of the full dynamics generated by $\mathcal L_0+\mathcal V_x$. The quantity $\langle\delta \hat n_x(t)\rangle=\mathrm{Tr}[\rho(t)(\hat n_x-\nu_s)]$ measures the deviation of the local occupation from this stationary filling.
Hence, Eq.~\eqref{EQ_U1ChargeVarianceExact} shows that the buildup of correlations between $\hat Q$ and $\hat n_x$ directly controls the charge-space broadening dynamics. 

The competition between $\mathcal L_0$ and $\mathcal{V}_x$ dynamics determines the scaling of correlation $C_x(t)$ and the local charge fluctuation $\langle \delta \hat{n}_x(t)\rangle$, which further determines the dynamical scaling of charge variance $\sigma^2(t)$. 
When bulk relaxation generated by $\mathcal L_0$ is fast compared with the local symmetry-charge exchange, the intra-sector state $\rho_Q(t)$ remains close to its steady state, while $C_x(t)$ and $\langle\delta\hat n_x(t)\rangle$ are approximately time independent, yielding charge-space diffusion, $\sigma_Q^2(t)\propto t$, according to Eq.~\eqref{EQ_U1ChargeVarianceExact}.
At sufficiently long times, however, the relaxation of the local disturbance is controlled by bulk hydrodynamics. 
For a dynamical exponent $z$, a local charge perturbation spreads over a length $\ell(t)\sim t^{1/z}$. 
For example, conventional diffusive $U(1)$ hydrodynamics corresponds to $z=2$~\cite{WangZhong2025PRB}. 
In one dimension, hydrodynamic memory causes $\chi-C_x(t)$ and $\langle\delta\hat n_x(t)\rangle$ to decay as $t^{-(1-1/z)}$, and Eq.~\eqref{EQ_U1ChargeVarianceExact} consequently yields anomalous charge-space broadening, $\sigma_Q^2(t)\propto t^{1/z}$. 

Thus, in the thermodynamic limit, the charge variance exhibits a two-stage algebraic scaling (see full derivations in Sup.~Mat.~B~\cite{SupMat}), 
\begin{equation}\label{EQ_U1ChargeVarianceScaling}
	\sigma_Q^2(t)\sim
	\begin{cases}
    [\kappa_+ (1-\nu_0) + \kappa_- \nu_0] t, & 0\ll t\ll t^*,\\
    \frac{z^2}{4\kappa \, \Gamma(1/z) \csc(\pi/z)}(Dt)^{1/z}, & t^* \ll t,
    \end{cases}
\end{equation}
where $t^* \sim (D/\kappa^z)^{1/(z-1)}$ denotes the crossover timescale between the Markovian charge-space broadening and the hydrodynamic-memory regime, $\nu_0$ represents the initial filling, and $\Gamma(1/z)$ is Gamma function. 
For conventional $U(1)$ hydrodynamics diffusion ($z=2$), $t^* \sim D/\kappa^2$, where $D$ is the bulk diffusion constant set by $\mathcal L_0$, and $\kappa$ characterizes the strength of the local symmetry-breaking channel given by $\mathcal V_x$. 

\begin{figure}[t]
	\begin{centering}
    \includegraphics[width=1\linewidth]{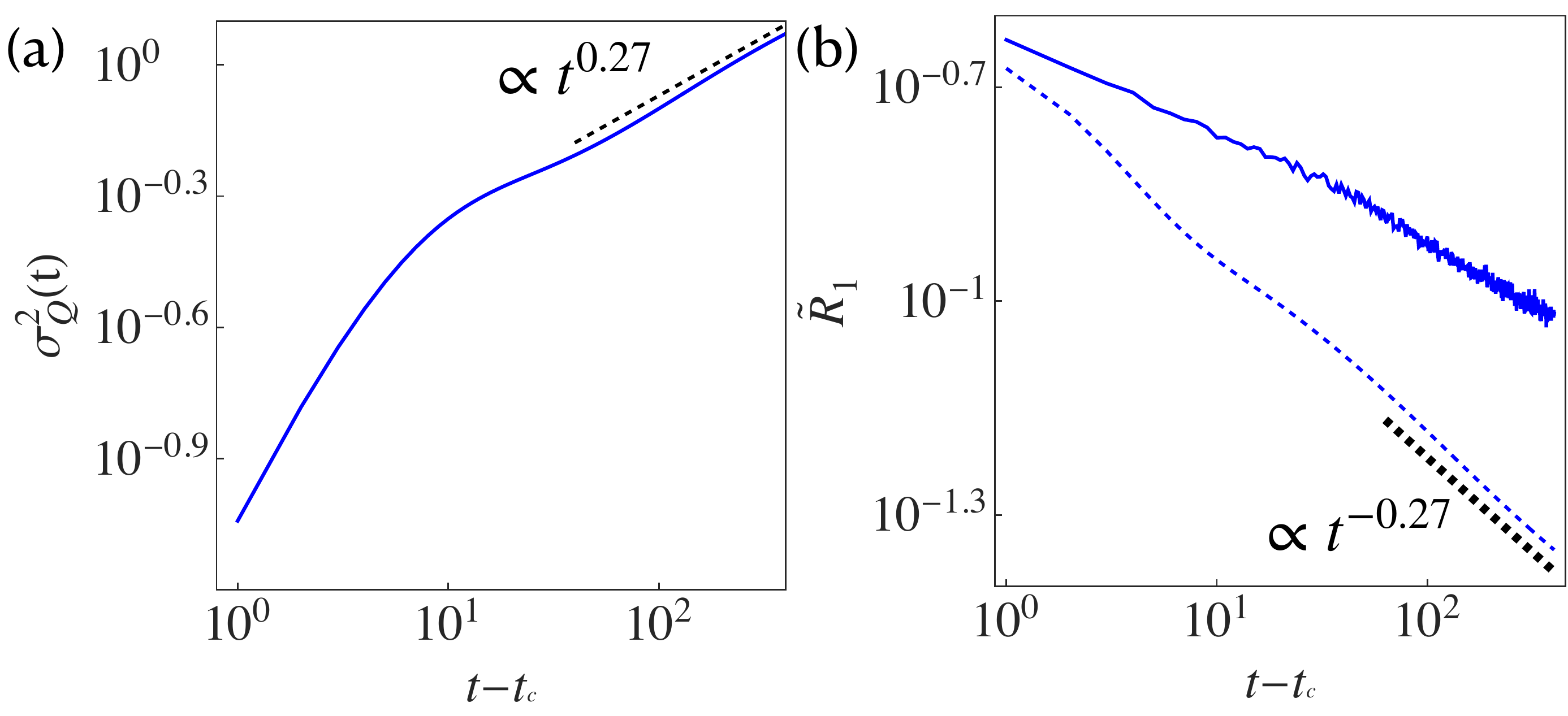}
    \par\end{centering}
	\protect\caption{\label{fig:3}Numerical results for the dipole-conserving model in Eq.~\eqref{EQ_simulation_model} with $L=50$, $\gamma_{\mathrm{hop}}=0.01$, $\gamma_\phi=5$, and $\kappa_+=\kappa_-=\kappa=0.1$, initialized in $\bar P_{L/2}$. (a) Charge variance $\sigma_Q^2(t)\propto(t-t_c)^{0.27}$, close to the predicted exponent $1/z=1/4$. (b) Log-log plot of the connected R\'enyi-1 correlator for $i=L/4$ and $j=3L/4$. The solid curve is the numerical result, while the blue dashed curve is the auxiliary-state prediction computed from the same $w_Q(t)$. Black dashed lines indicate the algebraic behaviors $(t-t_c)^{0.27}$ and $(t-t_c)^{-0.27}$ in (a) and (b), respectively.}
\end{figure}

Finally, combining Eqs.~\eqref{EQ_R1mms}-\eqref{EQ_U1ChargeVarianceScaling}, the auxiliary-state R\'enyi-1 correlator decays algebraically as
\begin{equation}\label{EQ_R1andSigmaQ}
	\lim_{|i-j|\to \infty} \widetilde R_1(\rho_{\mathrm{aux}}(t),c_i^\dagger c_j) \sim \frac{1}{\sigma_Q^2(t)}\sim t^{-\alpha}.
\end{equation}
Provided that $\overline{\mathcal F^2}(t)$ is lower bounded by a nonzero constant, Eq.~\eqref{EQ_R1R1MMS} then implies that the actual long-range R\'enyi-1 correlator cannot decay faster than algebraically. 
This algebraic robustness is in sharp contrast to the exponential fragility in the $\mathbb Z_2$ case. 

To verify these dynamical scalings numerically, we consider a fermion chain governed by
\begin{equation}\label{EQ_simulation_model}
	\mathcal L_0=\gamma_{\mathrm{hop}}\sum_j\left(\mathcal D[L_j]+\mathcal D[L_j^\dagger]\right)+\gamma_\phi\sum_{j=1}^{L}\mathcal D[n_j],
\end{equation}
with two choices of $L_j$ realizing distinct bulk dynamical exponents $z=2$ and $z=4$. 
For $L_j=c_{j+1}^\dagger c_j$, charge transport is diffusive at long times, with $z=2$. 
The initial state is MMS in the half-filled charge sector, $\rho(0)=\bar P_{L/2}$, we observe a progressive broadening of the sector-weight distribution $w_Q(t)$ [Fig.~\ref{fig:2}(a)]. 
At fixed $\gamma_\phi=5$ and $\kappa=0.1$, varying $\gamma_{\mathrm{hop}}$ tunes the crossover time $t^*\sim\gamma_{\mathrm{hop}}/\kappa^2$.
For fast hopping ($\gamma_{\mathrm{hop}}=2$), the large $t^*$ yields an extended regime of nearly linear charge-sector broadening, $\sigma_Q^2\propto(t-t_c)^{0.98}$ [red line in Fig.~\ref{fig:2}(b)]. 
By contrast, for slow hopping ($\gamma_{\mathrm{hop}}=0.01$), $t^*$ is much shorter, and the dynamics rapidly crosses over to the hydrodynamic-memory regime, where $\sigma_Q^2\propto(t-t_c)^{0.48}$ [blue line in Fig.~\ref{fig:2}(b)], consistent with the asymptotic prediction $\sigma_Q^2\sim (t-t_c)^{1/2}$. 
The R\'enyi-1 long-range correlation exhibits the corresponding algebraic crossover [Figs.~\ref{fig:2}(c)(d)] and is bounded by the analytically obtained auxiliary-state R\'enyi-1 correlations (the dashed lines). 
For $L_j=c_{j+3}^\dagger c_{j+2}c_{j+1}c_j^\dagger$, $\mathcal L_0$ conserves both the total charge and its dipole moment, giving rise to subdiffusive bulk dynamics with $z=4$ at long times. 
Figs.~\ref{fig:3}(a) and (b) show the corresponding algebraic charge-sector broadening and R\'enyi-1 decay, again bounded by the auxiliary-state result. 
Together, the diffusive and subdiffusive models demonstrate that the asymptotic relaxation of nonlinear long-range order follows $\widetilde R_1\sim t^{-1/z}$, with the exponent $z$ set by the underlying bulk hydrodynamics. 

\emph{{\clr Conclusion and Discussion.---}}
In summary, we reveal how the connected R\'enyi-1 long-range correlation decays once the strong symmetry is locally reduced to its weak counterpart.
For local strong-symmetry-breaking dynamics with $U(1)$ symmetry, we uncover parametrically slow algebraic relaxation and identify charge-sector broadening as its underlying mechanism. This behavior stands in sharp contrast to the exponential decay of the connected R\'enyi-1 correlation in the discrete $\mathbb{Z}_2$ symmetry class. 

The local symmetry-breaking channel broadens the distribution over symmetry-charge sectors. 
At short times, the disturbance is rapidly smeared out by the $\mathcal{L}_0$ dynamics, such that successive local charge-exchange processes effectively lose memory of one another and the charge-sector distribution broadens as a random walk in charge space. 
At long times, by contrast, the disturbance generated by $\mathcal{V}$ spreads through the lattice via hydrodynamic transport. Consequently, the charge-sector broadening inherits the hydrodynamic exponent and, in turn, determines the algebraic relaxation of the connected R\'enyi-1 long-range correlation. 
This charge-broadening mechanism provides a universal picture for the algebraic relaxation of SW-SSB order and can be straightforwardly generalized to arbitrary spatial dimensions. 
Furthermore, these dynamical scalings and the charge-broadening mechanism provide direct predictions that can be tested in realistic experimental settings, such as the dephased Fermi gas~\cite{IBloch2026}. 

\emph{{\clr Acknowledgement}.---}
The authors thank Chen Fang, Xiao-Qi Sun, Kai Sun and Chang Shu for valuable discussions. 
K.Z. acknowledges support by the start-up grant at Zhejiang University. 
Z.-X. F. is supported by the National Natural Science Foundation of China (Grant Nos. 12325404, 12547112, and 12188101), the National Key R\&D Program of China (Grant Nos. 2022YFA1403800 and 2023YFA1406704), and the Chinese Academy of Sciences Strategic Priority Research Program (B) (Grant No. XDB1720000).

\let\oldaddcontentsline\addcontentsline
\renewcommand{\addcontentsline}[3]{}

\let\addcontentsline\oldaddcontentsline
\onecolumngrid
\newpage
\makeatletter
\renewcommand \thesection{S-\@arabic\c@section}
\renewcommand\thetable{S\@arabic\c@table}
\renewcommand \thefigure{S\@arabic\c@figure}
\renewcommand \theequation{S\@arabic\c@equation}
\makeatother
\setcounter{equation}{0}  
\setcounter{figure}{0}  
\setcounter{section}{0}  
{\begin{center}
		{\bf \large Supplementary Materials for ``Algebraic Relaxation of Strong-to-Weak Symmetry-Breaking Order via Charge-Sector Broadening" }
\end{center}}
\maketitle
\tableofcontents

\section{Derivation of the dynamical equation for $w_Q(t)$}
In this section, we derive the dynamics of the symmetry-charge-sector weights $w_Q(t)$ under the strong-symmetry-breaking dynamics $\mathcal{L}_0 + \mathcal{V}_x$, i.e., Eq.~(10) in the main text.  
Since the channel $\mathcal{V}_x$ explicitly reduces the strong symmetry to weak $U(1)$ symmetry, so that the state generally takes a block-diagonal form in the charge sectors
\begin{equation}
	\rho(t)=\sum_Qw_Q(t)\rho_Q(t),\qquad \Tr[\rho_Q(t)]=1,\qquad \sum_Qw_Q(t)=1,
\end{equation}
where $w_Q(t)=\Tr[P_Q\rho(t)]$ and $\rho_Q(t)=P_Q\rho(t)P_Q/w_Q(t)$. 
Here, $P_Q$ denotes the projector onto the total $U(1)$ symmetry charge sector $Q$. 
Therefore, the time evolution of charge-sector weight $w_Q(t)$ becomes 
\begin{equation}\label{EQ_WeightTimeEvolve}
	\frac{dw_Q(t)}{dt} = \Tr\!\left[P_Q\dot\rho(t)\right]
	=\Tr\!\left[P_Q(\mathcal L_0+\mathcal V_x)[\rho(t)]\right].
\end{equation}
Since $\mathcal{L}_0$ has strong $U(1)$ symmetry, each charge-sector projector is invariant under its adjoint evolution, i.e., $\mathcal{L}^\dagger_0[P_Q] = 0$. 
Thus, 
\begin{equation*}
	\Tr[P_Q\mathcal L_0[\rho(t)]]=\Tr[\mathcal L_0^\dagger[P_Q]\rho(t)]=0, 
\end{equation*}
so the time evolution of $w_Q(t)$ is determined entirely by $\mathcal{V}_x$, as determined by Eq.~\eqref{EQ_WeightTimeEvolve}. 

For illustration, we take the local strong-symmetry-breaking channel to be
\begin{equation}\label{EQ_U1SymmBreaking}
	\mathcal V_x[\rho(t)] = \kappa_- \mathcal{D}[c_x] \rho(t) + \kappa_+ \mathcal{D}[c_x^\dagger] \rho(t),
\end{equation}
where $\mathcal D[O]\rho=O\rho O^\dagger-\frac12\{O^\dagger O,\rho\}$, with $\kappa_-$ and $\kappa_+$ the particle-loss and particle-gain rates, respectively. 
As illustrated in Fig.~\ref{subfig:1}, symmetry-charge exchange occurs only locally at site $x$, or more generally within a finite region in the thermodynamic limit.
We define charge-resolved particle and hole occupations as 
\begin{equation*}
	n_x^Q \equiv \Tr[n_x \rho_Q(t)]; \quad 
	\bar n_x^Q(t) \equiv 1 - n_x^Q(t).
\end{equation*}
Using $P_Qc_x=c_xP_{Q+1}$ and $P_Qc_x^\dagger=c_x^\dagger P_{Q-1}$, Eq.~\eqref{EQ_WeightTimeEvolve} becomes
\begin{equation}\label{eq:charge-sector-TE}
	\begin{split}
		\frac{dw_Q(t)}{dt} & = \Tr\!\left[P_Q\mathcal V_x[\rho(t)]\right] \\
		& = a_{Q+1}(t)w_{Q+1}(t)+b_{Q-1}(t)w_{Q-1}(t)-\left[a_Q(t)+b_Q(t)\right]w_Q(t),
	\end{split}
\end{equation}
where $a_Q(t)\equiv\kappa_-n_x^Q(t)$ and $b_Q(t)\equiv\kappa_+\bar n_x^Q(t)$. 
Finally, the dynamics of $w_Q(t)$ is fully determined by this equation together with the charge-space boundary conditions $a_0(t)=0$ and $b_L(t)=0$. 
This reproduces Eq.~(10) of the main text.

\section{Dynamical Exponent of the Charge Variance}

In this section, we derive the asymptotic scaling of the charge variance $\sigma_Q^2(t)\equiv\langle\hat Q^2\rangle-\langle\hat Q\rangle^2$ under both symmetry-breaking channel $\mathcal{V}_x$ and bulk dynamics $\mathcal{L}_0$. 

\subsection{Exact evolution of the charge variance}
We now derive the time evolution of charge variance  $\sigma_Q^2(t)\equiv\langle\hat Q^2\rangle-\langle\hat Q\rangle^2$, where $\hat Q=\sum_i \hat n_i$ is the total charge operator. 
To determine the time evolution of $\hat Q^m$ under Liouvillian dynamics, we use the Heisenberg picture, which gives 
\begin{equation}
	\frac{d \hat Q^m}{dt} = \mathcal{L}^\dagger[\hat Q^m] = \mathcal{L}_0^\dagger[\hat Q^m] + \mathcal{V}_x^\dagger[\hat Q^m],
\end{equation}
where $\mathcal{L}^\dagger$ denotes the adjoint Liouvillian superoperator. 
Since $\mathcal{L}_0$ preserves the strong U(1) symmetry, $\mathcal{L}_0^\dagger[\hat Q^m]=0$ and time evolution is determined by $\mathcal{V}_x^\dagger[\hat Q^m]$, i.e.,
\begin{equation}\label{EQ_mthMomentTE}
	\frac{d \hat Q^m}{dt} = \mathcal{V}_x^\dagger[\hat Q^m] = \kappa_- \mathcal{D}^{\dagger}[c_x] \hat Q^m + \kappa_+ \mathcal{D}^{\dagger}[c^{\dagger}_x] \hat Q^m,
\end{equation}
where $\mathcal{D}^{\dagger}[O]\hat Q^m = O^{\dagger}\hat Q^m O -\frac{1}{2}\{O^\dagger O, \hat Q^m\}$. 
Using commutation relation $[\hat Q, c_x] = -c_x$ and $[\hat Q, c_x^\dagger] = c_x^\dagger$, Eq.~\eqref{EQ_mthMomentTE} can be derived as 
\begin{equation}
	\begin{split}
		& \frac{d \hat Q}{dt} = - \kappa_- \hat n_x + \kappa_+ (1-\hat n_x); \\
		& \frac{d \hat Q^2}{dt} = \kappa_- (1-2 \hat Q) \hat n_x + \kappa_+ (1+2 \hat Q) (1-\hat n_x). 
	\end{split}
\end{equation}
Hence, the time evolution of charge variance can be obtained
\begin{equation}\label{eq:charge-variance-exact}
	\frac{d \sigma_Q^2}{dt} = \frac{d\langle\hat Q^2\rangle}{dt}-2 \langle\hat Q\rangle \frac{d\langle\hat Q\rangle }{dt} = \kappa_+ + (\kappa_--\kappa_+)\langle\hat n_x(t)\rangle - 2(\kappa_+ + \kappa_-) C_x(t),
\end{equation}
where $C_x(t)\equiv\operatorname{Cov}(\hat Q,\hat n_x)=\langle\hat Q\hat n_x\rangle-\langle\hat Q\rangle\langle\hat n_x\rangle$ is the covariance between $n_x$ and total charge $Q$. 
In what follows, we derive the dynamics of $\langle\hat n_x(t)\rangle$ and $C_x(t)$, from which we finally obtain the asymptotic scaling of the charge variance $\sigma_Q^2(t)$.

\subsection{Dynamics of the local occupation $\langle\hat n_x(t)\rangle$}
For convenience, we introduce the density fluctuation $\delta n_r = n_r - \nu_s$, where $\nu_s = \kappa_+/(\kappa_+ + \kappa_-)$ is the stationary local filling determined by the reservoir and $n_r$ is the occupation at site $r$. 
Taking Fourier transform, we have 
\begin{equation}
	\delta n_{k}(t) = \sum_r e^{-ik(r-x)}\delta n_r(t), 
\end{equation}
which measures the amplitude of a density fluctuation with wavevector $k$ relative to the symmetry-breaking site $x$. 
Using Heisenberg picture, the time evolution of the density fluctuation operator is determined by the adjoint Liouvillian superoperator, i.e., 
\begin{equation}\label{EQ_ChargeFluctationTE}
	\frac{d \, \delta n_k(t)}{dt} = \mathcal{L}_0^{\dagger}[\delta n_k(t)] + \mathcal{V}^\dagger_x [\delta n_k(t)].
\end{equation}
Our focus is on the long-time scaling behavior, where short-wavelength modes have already relaxed and the dynamics is dominated by long-wavelength physics. 
The evolution of long-wavelength charge fluctuations can therefore be approximated by the low-lying spectrum of the unperturbed Liouvillian, 
\begin{equation}
	\mathcal{L}_0^\dagger[\delta n_k(t)] \approx \lambda(k) \delta n_k(t); \quad \lambda(k) = -D|k|^z,
\end{equation}
where the dynamical exponent $z$ is determined by the underlying hydrodynamics associated with the choice of $\mathcal{L}_0$~\cite{dipole_Hydrodynamics_PRR_2020,Dynamical_univer_class_PNAS_2015}, and $D$ is a dynamical coefficient set by system parameters. 
The relaxation time of a mode with momentum $k$ scales as $\tau_k\sim 1/(D|k|^z)$, thus the small $k$ modes relax most slowly and dominate at a long-time scale. 
For the second term in Eq.~\eqref{EQ_ChargeFluctationTE}, it's straightforward to obtain 
\begin{equation}
	\mathcal{V}_x^\dagger[\delta n_k(t)] = \sum_r e^{-ik(r-x)} \mathcal{V}_x^\dagger[\delta n_r(t)] = - (\kappa_+ + \kappa_-) \delta n_x(t),
\end{equation}
which is similar to Eq.~\eqref{EQ_mthMomentTE}. 
Finally, we collect the above derivations and obtain the time evolution of the averaged density fluctuation
\begin{equation}\label{EQ_AveChargeFluctationTE}
	\frac{d \langle\delta n_k(t)\rangle}{dt} = \lambda(k) \langle \delta n_k(t)\rangle + S_0(t); \quad S_0(t) = -2 \kappa \,\langle\delta n_x(t)\rangle,
\end{equation}
with $\kappa \equiv (\kappa_+ + \kappa_-)/2$. 

The formal solution of Eq.~\eqref{EQ_AveChargeFluctationTE} is 
\begin{equation}
	\begin{aligned}
		\langle \delta n_k(t) \rangle=e^{\lambda(k)t}\langle\delta n_k(0)\rangle-2\kappa\int_0^t dt'\,e^{\lambda(k)(t-t')}\langle\delta n_x(t')\rangle.
	\end{aligned}
\end{equation}
Using the inverse Fourier transform and taking the continuum limit, we obtain $\langle \delta n_r(t)\rangle=\int_{-\Lambda}^{\Lambda}\frac{dk}{2\Lambda} e^{ik(r-x)} \langle \delta n_k(t)\rangle$, where $\Lambda$ is a long wavelength cutoff chosen such that higher momentum modes, which relax rapidly at long times, are neglected. 
Evaluating this expression at the symmetry-breaking site $r=x$, we obtain 
\begin{equation}\label{EQ_DensityFlucTE}
	\langle\delta n_x(t) \rangle = \int_{-\Lambda}^{\Lambda}\frac{dk}{2\Lambda} \langle \delta n_k(t)\rangle =\langle\delta n_x^{(0)}(t)\rangle-2\kappa\int_0^t dt'\,G_0(t-t')\langle\delta n_x(t')\rangle,
\end{equation}
where $\langle\delta n_x^{(0)}(t)\rangle=\int_{-\Lambda}^{\Lambda}\frac{dk}{2\Lambda}e^{\lambda(k)t}\langle\delta n_k(0)\rangle$, and $G_0(t)=\int_{-\Lambda}^{\Lambda}\frac{dk}{2\Lambda}e^{\lambda(k)t}$. 

In our consideration, the initial state is the maximally mixed state that has definite symmetry charge $Q_0$ and uniformly distribute over the lattice site. 
Therefore, there's no density fluctuation for the initial state, which means that $\langle\delta n_k(0)\rangle = (\nu_0-\nu_s) \delta_{k,0}$ and $\nu_0=Q_0/L$ is the initial filling. 
Plugging it into Eq.~\eqref{EQ_DensityFlucTE}, we have 
\begin{equation*}
	\langle \delta n_x^{(0)}(t)\rangle = \int_{-\Lambda}^{\Lambda}dk \, e^{\lambda(k)t} \, (\nu_0-\nu) \, \delta_{k,0} = \nu_0-\nu,
\end{equation*}
where $\lambda(k=0)=0$ has been used. 

We now apply Fourier transform, $f^R(\omega)=\int_0^\infty dt\,e^{i(\omega+i0^+)t}f(t)$, onto Eq.~\eqref{EQ_DensityFlucTE}, which gives $\langle\delta n^R_x(\omega) \rangle = \langle \delta n_x^{(0),R}(\omega)\rangle - 2\kappa G_0^R(\omega)\langle\delta n^R_x(\omega) \rangle$, and thus, $$\langle\delta n^R_x(\omega) \rangle = \frac{\langle \delta n_x^{(0),R}(\omega)\rangle}{1+2\kappa G_0^R(\omega)},$$ 
where $\langle\delta n_x^{(0),R}(\omega)\rangle=(\nu_0-\nu)/[-i(\omega+i0^+)]$. 
The retarded Green's function can be derived as 
\begin{equation}\label{EQ_RetardBareGreens}
	G_0^R(\omega)=\int_{-\Lambda}^{\Lambda}\frac{dk}{2\pi}\frac{1}{-i(\omega+i0^+)+D|k|^z}\simeq\frac{\csc(\pi/z)}{zD^{1/z}}\left[-i(\omega+i0^+)\right]^{1/z-1},
\end{equation}
where the cutoff works well for low-energy physics $|\omega|\ll D\Lambda^z$ and $z>1$.  
It follows that 
\begin{equation}\label{EQ_RetardPartFluc}
	\langle\delta n_x^R(\omega)\rangle\simeq(\nu_0-\nu_s)\frac{\left[-i(\omega+i0^+)\right]^{\alpha-1}}{\left[-i(\omega+i0^+)\right]^\alpha+B_0},\qquad \alpha=1-\frac{1}{z},\qquad B_0=\frac{2\kappa}{zD^{1/z}}\csc\left(\frac{\pi}{z}\right).
\end{equation}
Finally, the time evolution of the density fluctuation can be solved as 
\begin{equation}\label{EQ_LocalFillingTE}
	\begin{aligned}
		\langle\delta n_x(t)\rangle
		&=\int_{-\infty}^{\infty}\frac{d\omega}{2\pi}\,e^{-i\omega t} \langle\delta n_x^R(\omega)\rangle\\
		&=(\nu_0-\nu_s)\int_{-\infty}^{\infty}\frac{d\omega}{2\pi}\,e^{-i\omega t}\frac{\left[-i(\omega+i0^+)\right]^{\alpha-1}}{\left[-i(\omega+i0^+)\right]^\alpha+B_0}\\
		&=(\nu_0-\nu_s)\sum_{m=0}^{\infty}(-B_0)^m\int_{-\infty}^{\infty}\frac{d\omega}{2\pi}\,e^{-i\omega t}\left[-i(\omega+i0^+)\right]^{-(\alpha m+1)}\\
		&=(\nu_0-\nu_s)\sum_{m=0}^{\infty}\frac{(-B_0)^m}{\Gamma(\alpha m+1)}\int_0^\infty du\,u^{\alpha m}e^{-0^+u}\int_{-\infty}^{\infty}\frac{d\omega}{2\pi}\,e^{-i\omega(t-u)}\\
		&=(\nu_0-\nu_s)\sum_{m=0}^{\infty}\frac{(-B_0)^m}{\Gamma(\alpha m+1)}\int_0^\infty du\,u^{\alpha m}\delta(t-u)\\
		&=(\nu_0-\nu_s)\sum_{m=0}^{\infty}\frac{(-B_0t^\alpha)^m}{\Gamma(\alpha m+1)}\\
		&=(\nu_0-\nu_s)E_\alpha(-B_0t^\alpha).
	\end{aligned}
\end{equation}
Here $E_\alpha(y)=\sum_{m=0}^{\infty}y^m/\Gamma(\alpha m+1)$ is the Mittag-Leffler function~\cite{Gorenflo2020Book}.  
From the third to the fourth line in the above derivation, we used the integral representation $\left[-i(\omega+i0^+)\right]^{-\beta}=\Gamma(\beta)^{-1}\int_0^\infty du\,u^{\beta-1}e^{i(\omega+i0^+)u}$, which follows from the definition of the Gamma function $\Gamma(\beta)=\int_0^\infty dy\,y^{\beta-1}e^{-y}$. 
Finally, using $\langle\delta n_x(t)\rangle=\langle n_x(t)\rangle-\nu_s$, the local filling evolves as
\begin{equation}\label{EQ_LocalFillingDynamics}
	\langle n_x(t) \rangle = \nu_s + (\nu_0 - \nu_s) E_\alpha(-B_0t^\alpha); \quad \alpha = 1-\frac{1}{z},
\end{equation}
where $z$ is the dynamical exponent characterizing the underlying hydrodynamics. 
For example, $z=2$ for conventional diffusive transport, while $z=4$ for dipole-conserving hydrodynamics, as realized by the model in Fig.~3 of the main text. 
Therefore, Eq.~\eqref{EQ_LocalFillingDynamics} shows that the local filling relaxes toward the stationary value $\nu_s$ according to a Mittag-Leffler function, with the relaxation exponent set by the bulk hydrodynamics.

\subsection{Dynamics of the covariance $\operatorname{Cov}(\hat Q,\hat n_x)$}
We now derive the dynamics of covariance $C_x(t) \equiv \operatorname{Cov}(\hat Q,\hat n_x) = \langle \hat Q(t) \hat n_x(t)\rangle - \langle \hat Q(t) \rangle \langle \hat n_x(t)\rangle=\operatorname{Cov}(\delta \hat Q,\delta\hat n_x)$ with $\delta \hat Q=\hat Q-\nu_s L$ and $\delta \hat n_x=\hat n_x-\nu_s$.
Thus, the covariance $C_x(t) = \operatorname{Cov}(\delta \hat Q,\delta\hat n_x)$ measures how strongly a fluctuation of the global charge is correlated with the local occupation at the symmetry-breaking site $x$. 

Similar to the process to calculate $\langle n_x(t) \rangle$, we take the Fourier transform $C_k(t) = \sum_r e^{-ik(r-x)}C_r(t) = \operatorname{Cov}(\delta \hat Q,\delta\hat n_k)$.
The dynamics of $C_k(t)$ is determined by the adjoint Liouvillian superoperator 
\begin{equation}
	\begin{aligned}
		\frac{d C_k(t)}{dt} &=\langle\mathcal{L}^\dagger[\delta \hat Q \delta \hat{n}_k] \rangle-\langle \mathcal{L}^\dagger[\delta \hat Q] \rangle\langle \delta \hat{n}_k\rangle-\langle \delta \hat{Q}\rangle\langle \mathcal{L}^\dagger[\delta \hat{n}_k] \rangle\\
		&=\langle \delta \hat{Q} \mathcal{L}_0^\dagger [\delta\hat{n}_k]\rangle+\langle \mathcal{V}_x^\dagger[\delta \hat{Q} \delta \hat{n}_k] \rangle-\langle\mathcal{V}_x^\dagger[\delta\hat Q] \rangle\langle\delta \hat{n}_k \rangle-\langle\delta \hat{Q} \rangle\langle\mathcal{L}_0^\dagger[\delta\hat{n}_k] \rangle-\langle\delta\hat{Q} \rangle\langle\mathcal{V}_x^\dagger[\delta \hat{n}_k] \rangle\\
		&=\left(\langle \delta \hat{Q} \mathcal{L}_0^\dagger [\delta\hat{n}_k]\rangle-\langle\delta \hat{Q} \rangle\langle\mathcal{L}_0^\dagger[\delta\hat{n}_k] \rangle\right)
		+\left(\langle \mathcal{V}_x^\dagger[\delta \hat{Q} \delta \hat{n}_k] \rangle-\langle\mathcal{V}_x^\dagger[\delta\hat Q] \rangle\langle\delta \hat{n}_k \rangle-\langle\delta\hat{Q} \rangle\langle\mathcal{V}_x^\dagger[\delta \hat{n}_k] \rangle\right)
	\end{aligned}
\end{equation}
with the property $\mathcal{L}_0^\dagger[\delta\hat n_k]\simeq\lambda(k) \delta\hat n_k$. 
Therefore, similar to Eq.~\eqref{EQ_AveChargeFluctationTE}, the covariance evolves as 
\begin{equation}\label{EQ_CovTE}
	\frac{dC_k(t)}{dt}=\lambda(k)C_k(t) + S_k(t), \quad \lambda(k)=-D|k|^z, 
\end{equation}
with a source $S_k(t)= \langle \mathcal{V}_x^\dagger[\delta \hat{Q} \delta \hat{n}_k] \rangle-\langle\mathcal{V}_x^\dagger[\delta\hat Q] \rangle\langle\delta \hat{n}_k \rangle-\langle\delta\hat{Q} \rangle\langle\mathcal{V}_x^\dagger[\delta \hat{n}_k] \rangle$. 
Using the relation $\mathcal{V}_x^\dagger[\delta\hat{Q}]=\mathcal{V}_x^\dagger[\delta\hat{n}_k]=\mathcal{V}_x^\dagger[\delta\hat{n}_x]=-2\kappa \delta \hat{n}_x$ and $\mathcal{V}_x^\dagger [\delta\hat{Q} \delta\hat{n}_k]=4\kappa\chi +(\kappa_--\kappa_+)\delta \hat{n}_x-2\kappa(\delta \hat{n}_x \delta\hat{Q}+\delta \hat n_x\delta \hat{n}_k )$, $S_k(t)$ becomes
\begin{equation}
	\begin{aligned}
		\label{eq_Sk_with_gk}
		S_k(t) &=4\kappa\chi + (\kappa_--\kappa_+)\langle\delta \hat n_x(t)\rangle -2\kappa C_x(t)
		-2\kappa g_{k}(t),
	\end{aligned}
\end{equation}
where $\kappa\equiv(\kappa_+ + \kappa_-)/2$, $\nu_s=\kappa_+/(2\kappa)$ is the stationary filling, $\chi=\nu_s(1-\nu_s)$, and
\begin{equation}
	g_{k}(t)=\operatorname{Cov}(\delta\hat{n}_x,\delta\hat{n}_k)=\sum_j e^{-ik(j-x)}\operatorname{Cov}(\delta \hat n_x,\delta \hat n_j).
\end{equation}
We assume that the density fluctuation correlation $\operatorname{Cov}(\delta \hat n_x,\delta \hat n_j)$ has a correlation length $\xi$,
\begin{equation}
	\operatorname{Cov}(\delta\hat n_x,\delta\hat n_j) = A \exp\left[-\frac{(j-x)^2}{\xi^2}\right],
\end{equation}
The amplitude $A(t)$ is fixed by the zero-momentum sum rule $g_{k=0}(t)=\sum_jg_{xj}(t)=\operatorname{Cov}(\delta\hat n_x,\delta\hat Q)=C_x(t)$. 
Evaluating this sum within this Gaussian approximation gives $A(t)=C_x(t)/[\sqrt{\pi}\xi(t)]$, and the corresponding momentum-space profile is $g_k(t)=C_x(t)e^{-k^2\xi^2(t)/4}$. 
At long times, the dynamics is dominated by long-wavelength modes near $k=0$, thus, 
short-wavelength modes have decayed, so the dynamics is dominated by modes near $k=0$. $e^{-k^2\xi^2(t)/4}\simeq1$ and $g_k(t)\simeq C_x(t)$. 
Under this long-wavelength approximation, Eq.~\eqref{eq_Sk_with_gk} reduces to
\begin{equation}
	\mathcal S_k(t)=4\kappa[\chi-C_x(t)]+(\kappa_--\kappa_+)\langle\delta\hat n_x(t)\rangle.
\end{equation}
The formal solution for $C_k(t)$ thus becomes 
\begin{equation}\label{eq:Ck-general}
	C_k(t)=e^{\lambda(k)t}C_k(0)+\int_0^t dt' e^{\lambda(k)(t-t')}
	\Big\{4\kappa[\chi-C_x(t')]+(\kappa_--\kappa_+)\langle\delta \hat n_x(t')\rangle\Big\}.
\end{equation}
After Fourier transform $C_k(t) \to C_r(t)$ and fix $r$ at the symmetry-breaking site $r=x$, we obtain
\begin{equation}\label{eq:Cx-general}
	\begin{split}
		C_x(t)&=C_x^{(0)}(t)+4\kappa\int_0^t dt'G_0(t-t')[\chi-C_x(t')]+(\kappa_--\kappa_+)\int_0^t dt'G_0(t-t')\langle\delta n_x(t')\rangle,
	\end{split}
\end{equation}
where $C_x^{(0)}(t)=\int_{-\Lambda}^{\Lambda}\frac{dk}{2\pi}\,e^{\lambda(k)t}C_k(0)$. 
For any initial state supported entirely in a fixed-charge sector $Q_0$, one has $C_r(0)=\operatorname{Cov}(\hat Q,\hat n_r)=0$, hence $C_k(0) = \sum_r e^{-ik(r-x)}C_r(0) =0$, therefore $C_x^{(0)}(t)=\int_{-\Lambda}^{\Lambda}\frac{dk}{2\pi}\,e^{\lambda(k)t}C_k(0)=0$.
Consequently, Eq.~\eqref{eq:Ck-general} and Eq.~\eqref{eq:Cx-general} reduce to 
\begin{equation}
	\begin{aligned}
		C_k(t)&=\int_0^t dt' e^{\lambda(k)(t-t')}
		\Big\{4\kappa[\chi-C_x(t')]+(\kappa_--\kappa_+)\langle\delta \hat n_x(t')\rangle\Big\},\\
		C_x(t)&=\int_0^t dt'\Big\{4\kappa G_0(t-t')[\chi-C_x(t')]+(\kappa_--\kappa_+)G_0(t-t')\langle\delta \hat n_x(t')\rangle\Big\}.
	\end{aligned}
	\label{eq:Ck-Cx-integral}
\end{equation}
Applying the Fourier transform $f^R(\omega)=\int_0^\infty dt\,e^{i(\omega+i0^+)t}f(t)$ to the second expression of Eq.~\eqref{eq:Ck-Cx-integral} gives
\begin{equation}
	C_x^R(\omega)=4\kappa G_0^R(\omega)\left[\frac{\chi}{-i(\omega+i0^+)}-C_x^R(\omega)\right]+(\kappa_--\kappa_+)G_0^R(\omega)\langle \delta\hat n_x^R(\omega)\rangle,
\end{equation}
and hence
\begin{equation}
	C_x^R(\omega)=\frac{4\kappa\chi G_0^R(\omega)}{-i(\omega+i0^+)\left[1+4\kappa G_0^R(\omega)\right]}+\frac{(\kappa_--\kappa_+)G_0^R(\omega)\langle\delta n_x^R(\omega)\rangle}{1+4\kappa G_0^R(\omega)}.
	\label{eq:Cx-frequency-general}
\end{equation}

Using $G_0^R(\omega)\simeq\csc(\pi/z)\left[-i(\omega+i0^+)\right]^{-\alpha}/(zD^{1/z})$, $\langle\delta n_x^R(\omega)\rangle\simeq(\nu_0-\nu)\left[-i(\omega+i0^+)\right]^{\alpha-1}/\{\left[-i(\omega+i0^+)\right]^\alpha+B_0\}$, and $B_0=2\kappa\csc(\pi/z)/(zD^{1/z})$ given by Eq.~\eqref{EQ_RetardBareGreens} and Eq.~\eqref{EQ_RetardPartFluc}, we can rearrange Eq.~\eqref{eq:Cx-frequency-general} as
\begin{equation}\label{eq:Cx-frequency-decomposed}
	\begin{aligned}
		C_x^R(\omega)\simeq{}&
		\chi\left[
		\frac{1}{-i(\omega+i0^+)}
		-\frac{\left[-i(\omega+i0^+)\right]^{\alpha-1}}
		{\left[-i(\omega+i0^+)\right]^\alpha+2B_0}
		\right]\\
		&+\frac{\kappa_--\kappa_+}{2\kappa}(\nu_0-\nu)
		\left[
		\frac{\left[-i(\omega+i0^+)\right]^{\alpha-1}}
		{\left[-i(\omega+i0^+)\right]^\alpha+B_0}
		-\frac{\left[-i(\omega+i0^+)\right]^{\alpha-1}}
		{\left[-i(\omega+i0^+)\right]^\alpha+2B_0}
		\right].
	\end{aligned}
\end{equation}
Using the inverse-Fourier-transform identity derived in Eq.~\eqref{EQ_LocalFillingTE},
$\mathcal F_R^{-1}\{\left[-i(\omega+i0^+)\right]^{\alpha-1}/[\left[-i(\omega+i0^+)\right]^\alpha+a]\}=E_\alpha(-a t^\alpha)$, we finally obtain
\begin{equation}\label{eq:Cx-general-solution}
	C_x(t)\simeq
	\chi\left[1-E_\alpha(-2B_0t^\alpha)\right]
	+\frac{\kappa_--\kappa_+}{2\kappa}(\nu_0-\nu_s)
	\left[E_\alpha(-B_0t^\alpha)-E_\alpha(-2B_0t^\alpha)\right].
\end{equation}
where $E_\alpha(y)=\sum_{n=0}^{\infty}y^n/\Gamma(\alpha n+1)$ is the Mittag--Leffler function, having a asymptotic form $E_\alpha(-z)\sim (z\Gamma(1-\alpha))^{-1}$ as $z$ approaches infinity.

\subsection{Asymptotic scaling of the charge variance}
It follows from Eqs.~\eqref{EQ_LocalFillingTE} and~\eqref{eq:Cx-general-solution} that the time evolution of the charge variance can be derived as
\begin{equation}\label{eq:charge-variance-general-ML}
	\begin{aligned}
		\frac{d\sigma_Q^2(t)}{dt}
		&=4\kappa[\chi-C_x(t)]+(\kappa_--\kappa_+)\langle\delta n_x(t)\rangle\\
		&\simeq
		4\kappa\chi E_\alpha(-2B_0t^\alpha)+(\kappa_--\kappa_+)(\nu_0-\nu_s)\left[2E_\alpha(-2B_0t^\alpha)-E_\alpha(-B_0t^\alpha)\right],
	\end{aligned}
\end{equation}
where $B_0=\frac{2\kappa}{zD^{1/z}}\csc\left(\frac{\pi}{z}\right)$, and $E_\alpha(-a t^{\alpha})$ is the Mittag--Leffler function that has the asymptotic form at short and late time scales: at short times $-a\,t^{\alpha} \ll 1$, $E_\alpha(-a t^{\alpha})\simeq 1$; while at long times $-a t^{\alpha} \gg 1$, $E_\alpha(-at^\alpha)\simeq[a\Gamma(1-\alpha)t^\alpha]^{-1}$.
Thus, the crossover time can be defined as 
\begin{equation}
	t^* \sim (B_0)^{-1/\alpha}=\left[\frac{zD^{1/z}}{2\kappa\csc(\pi/z)}\right]^{z/(z-1)}\sim\left(\frac{D}{\kappa^z}\right)^{1/(z-1)},
\end{equation}
or equivalently, 
\begin{equation}
	\kappa t^* \sim \left(\frac{D}{\kappa}\right)^{1/(z-1)}.
\end{equation}

Using the asymptotic form at short and late time scales of the Mittag--Leffler function, the time evolution of charge variance can be obtained in these two time scales as 
\begin{equation}
	\frac{d\sigma_Q^2(t)}{dt}\simeq
	\begin{cases}
		4\kappa\chi+(\kappa_--\kappa_+)(\nu_0-\nu),&t\ll t^*,\\
		\frac{zD^{1/z}}{4\kappa\Gamma(1/z)\csc(\pi/z)} t^{-(1-1/z)},&t\gg t^*,
	\end{cases}
	\label{eq:charge-variance-rate-asymptotic}
\end{equation}
where the short-time coefficient can equivalently be written as $\kappa_+(1-\nu_0)+\kappa_-\nu_0$. 
Integrating Eq.~\eqref{eq:charge-variance-rate-asymptotic} with initial condition $\sigma_Q^2(0)=0$ gives
\begin{equation}\label{eq:charge-variance-scaling}
	\sigma_Q^2(t)\sim
		\begin{cases}
			\left[\kappa_+(1-\nu_0)+\kappa_-\nu_0\right]t, & t\ll t^*,\\[1mm]
			\frac{ z^2}{4\kappa\Gamma(1/z)\csc(\pi/z)}(Dt)^{1/z}, &t\gg t^*.
	\end{cases}
\end{equation}
From Eq.~\eqref{eq:charge-variance-rate-asymptotic} and Eq.~\eqref{eq:charge-variance-scaling} reveal two distinct dynamical regimes.

At early times, $ t\ll t^*$, the bulk dynamics $\mathcal{L}_0$ rapidly relaxes the local disturbance generated by $\mathcal{V}_x$, such that successive charge-exchange events are effectively uncorrelated. 
The total charge therefore performs a random walk in charge-sector space, giving the normal-diffusive scaling $\sigma_Q^2(t)\sim t$. 

At late times, $ t\gg t^*$, charge injected or removed by $\mathcal{V}_x$ spreads into the bulk through the hydrodynamic modes of $\mathcal{L}_0$. The local symmetry-breaking channel thus retains a long-time hydrodynamic memory, leading to the anomalous scaling $\sigma_Q^2(t)\sim(Dt)^{1/z}$, with the dynamical exponent $z$ set solely by the underlying hydrodynamics. 
In the main text, we numerically verify these dynamical scalings using the representative cases $z=2$ and $z=4$ in Fig.~2 and Fig.~3, respectively. 
The former corresponds to conventional $U(1)$ hydrodynamics with diffusive charge transport, while the latter arises when an additional dipole conservation renders the charge dynamics subdiffusive.

\section{Lower bound for the decay of R\'enyi-k long-range correlation}\label{sec_MMSRenyi2}

In this section, we derive a lower bound for the R\'enyi-2 long-range correlation in the first three parts and generalize the result to R\'enyi-k correlations in the final part. 
We show that the connected R\'enyi-k long-range correlation of the evolving mixed state $\rho(t)$ is lower bounded by the corresponding correlation of an auxiliary state $\rho_{\text{aux}}(t)$, thereby reproducing the lower bound in Eq.~(8) of the main text (i.e., Eq.~\eqref{eq:k1-fidelity-lower-bound} in this section). 

Here we clarify the notation used in the following discussion.
Under the symmetry-breaking channel $\mathcal{V}$, the strong symmetry of the Liouvillian dynamics is broken down to the weak symmetry. 
The mixed state with weak U(1) symmetry can be written as $\rho(t) = \sum_Q w_Q(t) \rho_Q(t)$, with $w_Q(t)$ the charge-sector weight and $\rho_Q(t)$ the charge-sector state.
It satisfies the normalization condition $\Tr[\rho]=1$ and $\sum_Q w_Q = 1$, and thus $\Tr[\rho_Q]=1$. 
For the typical strongly U(1)-symmetric dissipative dynamics $\mathcal{L}_0$ in the presence of on-site dephasing, each fixed-charge sector is relaxed to the maximally mixed state (MMS) $\bar P_Q = P_Q/\Tr[P_Q]$. 
Therefore, it's convenient to write the charge-sector state as 
\begin{equation}\label{EQ_IntraSectorState}
	\rho_Q = \bar P_Q + \bar \delta_Q = (P_Q + \delta_Q)/D_Q; \quad D_Q = \Tr[P_Q],
\end{equation}
where $\delta_Q $ represents the deviation from the charge-sector MMS and is traceless $\Tr[\delta_Q]=0$. 

\subsection{Step 1: Bound for the single-point R\'enyi-2 correlation}

\begin{figure}[h]
	\begin{centering}
		\includegraphics[width=1\linewidth]{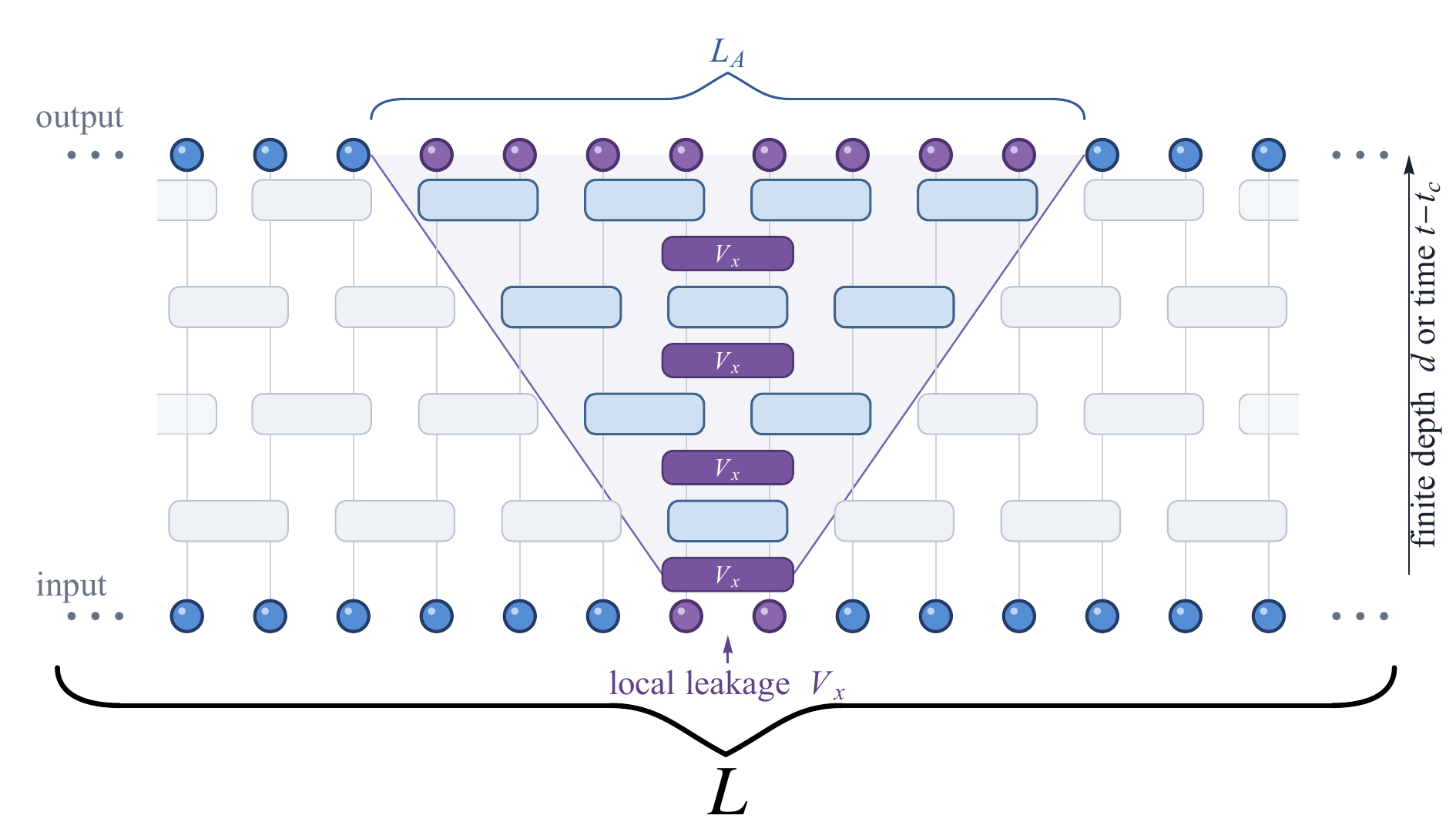}
		\par\end{centering}
	\protect\caption{\label{subfig:1}
		Schematic of the subsystem partition used in the derivation. The local leakage channel $V_x$ acts at the center in each circuit layer, while the remaining blocks represent local two-site channels. After a finite depth $d$, corresponding to an evolution time $t-t_c$, the shaded cone encloses the leakage-affected region $A$ of length $L_A$. Purple and blue sites denote $A$ and its complement $B$, respectively, and $L$ is the total system size. }
\end{figure}

As illustrated in Fig.~\ref{subfig:1}, the disturbance caused by the local symmetry-breaking channel spreads over a characteristic length $\ell_D(t)\sim\sqrt{D(t-t_c)}$ under unperturbed Liouvillian dynamics $\mathcal{L}_0$, where $D$ is the diffusion constant set by parameters in $\mathcal{L}_0$. 
Therefore, we bipartition the lattice into regions A and B, where A is chosen sufficiently large to contain the entire spreading region (the purple region in Fig.~\ref{subfig:1}), while B denotes the rest of the lattice, where the charge distribution has not yet been affected by the symmetry-breaking dynamics. 
Such a bipartition can always be defined because the dynamics is considered in the thermodynamic limit. 

We denote the lengths of the two regions by $L_A$ and $L_B$, respectively, with $L_A \sim \ell_D(t)$ and $L_A \ll L_B$. 
For region $X=A,B$, we denote the total symmetry charge in this region by $\hat Q_X=\sum_{i\in X} \hat n_i$. 
The initial state is chosen to be the MMS $\rho(0) = \bar P_{Q_0}$, which respects the strong $U(1)$ symmetry and exhibits the SW-SSB long-range order. 
The initial state can be decomposed into 
\begin{equation}
	\rho(0)=\bar P_{Q_0}=\frac{1}{D_{Q_0}}\sum_q P_A^q\otimes P_B^{Q_0-q}, 
\end{equation}
where $D_{Q_0}=\Tr[P_{Q_0}]$ is the normalization factor. 
As time evolves, the $\mathcal{L}_0$ dynamics only affects the distribution in the region A, hence, we can further represent the time-evolved mixed state as 
\begin{equation}\label{EQ_MixedStateAnsatz}
	\rho(t) = \sum_Q w_Q(t) \rho_Q(t); \quad 
	\rho_Q(t) = \bar P_Q + \bar \delta_Q(t) = \bar P_Q+\sum_q \rho_{A,Q}^q(t) \otimes \bar P_B^{Q-q}, 
\end{equation}
where $\rho_{A,Q}^q(t)$ is the deviation from MMS in region A associated with charge $q$ and also depends on the total charge $Q$. 

The one-point R\'enyi-2 correlation is defined as $R_2(\rho(t), c_i) = \Tr[\rho(t) c_i^{\dagger} \rho(t) c_i]/\mathcal Z_2$, where $\mathcal Z_2 = \Tr[\rho^2(t)]$ is the purity. 
We define the spatially averaged diagonal correlator for region B
\begin{equation}\label{eq:R2-spatial-sum}
	\mathcal R^{\rm diag}_{2}(\rho)\equiv\frac{1}{L_B}\sum_{i\in B}R_2(\rho,c_i)=\frac{\mathcal N_{B,\text{diag}}}{\mathcal Z_2}; \qquad \mathcal N_{B,\text{diag}}=\frac{1}{L_B}\sum_Qw_{Q+1}w_Q\sum_{i \in B}\Tr[\rho_{Q+1} c_i^\dagger \rho_Q c_i].
\end{equation}
In our setup, as shown in Fig.~\ref{subfig:1}, region A contains the region reached by the local symmetry-breaking disturbance, whereas B is the much larger, essentially unaffected region. 
The spatial average over region B therefore probes the overlap associated with moving or adding a particle at site $i$ far away from the symmetry-breaking region. 

Substituting Eq.~\eqref{EQ_IntraSectorState} into Eq.~\eqref{eq:R2-spatial-sum}, we have four terms
\begin{equation}\label{EQ_DiagonalParts}
	\begin{split}
		\mathcal N_{B,\text{diag}} & = \mathcal N_{B,\text{diag}}^{(0)} + \mathcal N_{B,\text{diag}}^{(1)} + \mathcal N_{B,\text{diag}}^{(2)} \\
		&= \sum_{Q} w_{Q+1}w_Q \frac{1}{L_B} \sum_{i\in B} \left( \Tr[\bar P_{Q+1} c_i^\dagger \bar P_{Q}c_i ] + \Tr[\bar P_{Q+1} c_i^\dagger \bar \delta_{Q}c_i ] + \Tr[\bar \delta_{Q+1} c_i^\dagger \bar P_{Q} c_i ] + \Tr[\bar \delta_{Q+1} c_i^\dagger \bar \delta_{Q} c_i ] \right),
	\end{split}
\end{equation}
where the superscript $(i)$ corresponds to $i$-th order of derivation term $\delta_Q$. 
It's directly calculated to be
\begin{equation}
	\mathcal N_{B,\text{diag}}^{(0)} = \sum_Q \frac{w_{Q+1}w_Q}{D_{Q+1}D_Q} \binom{L-1}{Q}, \qquad \mathcal N_{B,\text{diag}}^{(2)}=\frac{1}{L_B}\sum_Q\frac{w_{Q+1}w_Q}{D_{Q+1}D_Q}\sum_{i\in B}\Tr[\delta_{Q+1} c_i^\dagger \delta_Q c_i].
	\label{eq:R2-local-numerator-decomposition}
\end{equation}
For the linear-order term $\mathcal N_{B,\text{diag}}^{(1)}$, using the relation $\hat c_i^\dagger \hat P_Q \hat c_i = \hat n_i \hat P_{Q+1}$, we have 
\begin{equation*}
	\frac{1}{L_B}\sum_{i\in B}\Tr[\bar\delta_{Q+1}c_i^\dagger\bar P_Qc_i]=\frac{\Tr[\hat \delta_{Q+1} (\hat Q-\hat Q_A)]}{L_BD_QD_{Q+1}} = \frac{(Q+1)\Tr[\hat \delta_{Q+1}]- \Tr[\hat \delta_{Q+1} \hat Q_A]}{L_BD_QD_{Q+1}} = -\frac{\Tr[\hat \delta_{Q+1} \hat Q_A]}{L_BD_QD_{Q+1}},
\end{equation*}
where the traceless condition $\Tr[\hat \delta_{Q+1}]=0$ has been used in the last equality. 
Similarly, 
\begin{equation*}
	\frac{1}{L_B}\sum_{i\in B}\Tr[\bar P_{Q+1}c_i^\dagger\bar\delta_Qc_i]=\frac{\Tr[\hat Q_A \hat \delta_Q]}{L_BD_QD_{Q+1}}.
\end{equation*}
It follows that 
\begin{equation}
	\mathcal N_{B,\text{diag}}^{(1)} = \frac{1}{L_B} \sum_Q \frac{w_{Q+1}w_Q}{D_{Q+1}D_Q} \Tr[\hat Q_A (\hat \delta_{Q}-\hat \delta_{Q+1})]
\end{equation}
Using the traceless condition $\Tr[\hat \delta_{Q}]=0$ again, the purity in Eq.~\eqref{eq:R2-spatial-sum} can be decomposed as 
\begin{equation}\label{eq:purity-deviation}
	\mathcal Z_2=\mathcal Z_2^{(0)}+\mathcal Z_2^{(2)};\qquad \mathcal Z_2^{(0)}=\sum_Q\frac{w_Q^2}{D_Q},\qquad \mathcal Z_2^{(2)}=\sum_Q\frac{w_Q^2}{D_Q^2}\Tr[\delta_Q^2] \geq 0.
\end{equation}
Since $\|\hat Q_A\|_\infty\leq L_A$, the contribution of the linear term $\mathcal N_{B,\text{diag}}^{(1)}$ can be estimated in the order of
\begin{equation}
	\frac{\mathcal N_{B,\text{diag}}^{(1)}}{\mathcal{Z}_2} = \mathcal{O}\left(\frac{L_A}{L_B}\right). 
\end{equation}

We define the auxiliary mixed state 
\begin{equation}\label{EQ_AuxMixState}
	\rho_{\text{aux}}(t) = \sum_Q w_Q(t) \bar P_Q,
\end{equation}
where the charge-sector state $\rho_Q(t)$ is replaced by MMS $\bar P_Q$, such that the time evolution of the auxiliary mixed state is fully encoded in charge-sector distribution $w_Q(t)$. 
The corresponding averaged one-point R\'enyi-2 correlation for region B is simply 
\begin{equation}
	\mathcal{R}_{2,\text{aux}}^\text{diag} = \mathcal N_{B,\text{diag}}^{(0)}/\mathcal{Z}_2^{(0)}.
\end{equation}
Consequently, combining the above equations, the deviation from the auxiliary-state R\'enyi-2 correlation becomes
\begin{equation}\label{eq:R2-difference-mm}
	\mathcal R^{\text{diag}}_{2}-\mathcal{R}_{2,\text{aux}}^\text{diag}=\frac{\mathcal N_{B,\text{diag}}^{(2)}-\mathcal{R}_{2,\text{aux}}^\text{diag} \mathcal Z_2^{(2)}}{\mathcal Z_2}+\mathcal O\!\left(\frac{L_A}{L_B}\right).
\end{equation}
At any finite evolution time, the spreading region remains finite compared with the total system size. Therefore, $L_A/L_B\to 0$ in the thermodynamic limit. In this limit, the leading deviation from the auxiliary-state result is governed by terms quadratic in $\delta_Q$.

\subsection{Step 2: Bound for the off-diagonal R\'enyi-2 correlation}

We now consider the off-diagonal long-range R\'enyi-2 correlation $\lim_{|i-j|\to \infty} \mathcal{R}_{2,ij} = \Tr[\rho(t)c_i^\dagger c_j \rho(t) c_j^\dagger c_i]/\mathcal{Z}_2$. 
It's worth noting that in the thermodynamic limit, the long-range off-diagonal correlation $\lim_{|i-j|\to \infty} \mathcal{R}_{2,ij}$ can be equivalently characterized by the spatially averaged off-diagonal correlation over region B
\begin{equation}\label{eq:R2-long-range-definition}
	\mathcal R_{2}^\text{off} = \frac{\mathcal N_{B,\text{off}}}{\mathcal Z_2}, \qquad 
	\mathcal N_{B,\text{off}}=\frac{1}{L_B(L_B-1)}\sum_Qw_Q^2\sum_{\substack{i,j\in B\\i\neq j}}\Tr[\rho_Qc_i^\dagger c_j\rho_Qc_j^\dagger c_i].
\end{equation}
The number of pairs $(i,j)$ with distance $ l \lesssim L_A$ has the order of $\mathcal{O}(L_B l)$, therefore, their fraction among all pairs $\mathcal{O}(L_B^2)$ is $\mathcal{O}(L_B l)/\mathcal{O}(L_B^2) \sim \mathcal{O}(l/L_B)$, which vanishes in the thermodynamic limit. 
Therefore, in the thermodynamic limit, almost every pair of sites in $B$ is separated by a distance much larger than $L_A$. 
Therefore, the spatially averaged correlator over $B$ in Eq.~\eqref{eq:R2-long-range-definition} characterizes the long-range correlation. 

Similar to Eq.~\eqref{EQ_DiagonalParts}, the averaged off-diagonal correlation over region B can be decomposed into
\begin{equation}
	\mathcal{N}_{B,\text{off}} = \mathcal{N}_{B,\text{off}}^{(0)} + \mathcal{N}_{B,\text{off}}^{(1)} + \mathcal{N}_{B,\text{off}}^{(2)}, 
\end{equation}
with the superscript $(i)$ corresponding to the $i$-th order in $\delta_Q(t)$. 
For $i\neq j$, the operator $c_i^\dagger c_j$ preserves the total charge and hence commutes with $P_Q$. Moreover, $c_i^\dagger c_jc_j^\dagger c_i=n_i(1-n_j)$, which gives $\Tr[P_Qn_i(1-n_j)]=\binom{L-2}{Q-1}$. 
Therefore, the averaged off-diagonal correlation calculated from the auxiliary mixed state in Eq.~\eqref{EQ_AuxMixState} becomes 
\begin{equation}\label{eq:R2-long-range-mm}
	\mathcal R_{2,\text{aux}}^{\rm off} = \frac{\mathcal{N}_{B,\text{off}}^{(0)}}{\mathcal Z_2^{(0)}}; \quad 
	\mathcal{N}_{B,\text{off}}^{(0)} = \sum_Q\frac{w_Q^2}{D_Q^2}\binom{L-2}{Q-1}.
\end{equation}
Following the similar calculation, we can finally calculate the averaged off-diagonal correlation derivation from that calculated from the auxiliary mixed state as 
\begin{equation}\label{eq:R2-long-range-difference}
	\mathcal R_{2}^{\text{off}}-\mathcal R_{2,\text{aux}}^{\rm off} = \frac{\mathcal N_{B,\text{off}}^{(2)}-\mathcal R_{2,\text{aux}}^{\rm off}\mathcal Z_2^{(2)}}{\mathcal Z_2} + \mathcal{O}\!\left(\frac{L_A}{L_B}\right).
\end{equation}

\subsection{Step 3: Bound for the connected R\'enyi-2 correlator}
We now examine the connected R\'enyi-2 correlator. 
As discussed above, the long-range connected R\'enyi-2 correlation can be characterized by the connected spatially averaged correlator: $\widetilde{\mathcal R}_2=\mathcal R_2^{\text{off}}-(\mathcal R_2^{\text{diag}})^2$, with $\widetilde{\mathcal R}_{2,\text{aux}}=\mathcal R_{2,\text{aux}}^{\text{off}}-(\mathcal R_{2,\text{aux}}^{\text{diag}})^2$ for the auxiliary state. 
According to Eqs.~\eqref{eq:R2-difference-mm} and~\eqref{eq:R2-long-range-difference}, their difference is given by
\begin{equation}\label{eq:R2-connected-diff}
	\begin{aligned}
		\widetilde{\mathcal R}_2-\widetilde{\mathcal R}_{2,\text{aux}}
		={}&\frac{\epsilon_\delta^2}{1+\epsilon_\delta^2}
		\left[
		\frac{\mathcal N_{B,\text{off}}^{(2)}}{\mathcal Z_2^{(2)}}
		-\left(\frac{\mathcal N_{B,\text{diag}}^{(2)}}{\mathcal Z_2^{(2)}}\right)^2
		-\widetilde{\mathcal R}_{2,\text{aux}}
		\right]\\
		&+\frac{\epsilon_\delta^2}{(1+\epsilon_\delta^2)^2}
		\left(
		\frac{\mathcal N_{B,\text{diag}}^{(2)}}{\mathcal Z_2^{(2)}}
		-\mathcal R_{2,\text{aux}}^{\text{diag}}
		\right)^2\\
		&+\mathcal O\!\left(\frac{L_A}{L_B}\right).
	\end{aligned}
\end{equation}
Here, $\epsilon_\delta^2 = \mathcal Z_2^{(2)}/\mathcal Z_2^{(0)}$, and thus $1+\epsilon_\delta^2=\mathcal Z_2/\mathcal Z_2^{(0)}$ measures the purity of the mixed state relative to that of the auxiliary mixed state defined in Eq.~\eqref{EQ_AuxMixState}.  
All terms linear in $\delta_Q$ contribute only at order $\mathcal O(L_A/L_B)$, which vanishes in the thermodynamic limit for any finite evolution time, or equivalently for any finite quantum-circuit depth as illustrated in Fig.~\ref{subfig:1}. 
Therefore, the leading difference between the connected long-range R\'enyi-2 correlations of the mixed state and its auxiliary state is governed by terms quadratic in $\delta_Q$, namely, $\mathcal N_{B,\text{diag}}^{(2)}$ and $\mathcal N_{B,\text{off}}^{(2)}$, as shown in Eq.~\eqref{eq:R2-connected-diff}. 
Since the second line in Eq.~\eqref{eq:R2-connected-diff} is always positive and $\mathcal O(L_A/L_B)$ tends to vanishes in the thermodynamic limit, we can release Eq.~\eqref{eq:R2-connected-diff} into the inequality
\begin{equation}\label{eq:R2-diff-inequality}
	\begin{split}
		\widetilde{\mathcal R}_2-\widetilde{\mathcal R}_{2,\text{aux}}
		\geq{}&\frac{\epsilon_\delta^2}{1+\epsilon_\delta^2}
		\left[
		\frac{\mathcal N_{B,\text{off}}^{(2)}}{\mathcal Z_2^{(2)}}
		-\left(\frac{\mathcal N_{B,\text{diag}}^{(2)}}{\mathcal Z_2^{(2)}}\right)^2
		-\widetilde{\mathcal R}_{2,\text{aux}}
		\right].
	\end{split}
\end{equation}

As given in Eq.~\eqref{EQ_MixedStateAnsatz}, the charge-sector deviation can be expressed as $\bar\delta_Q(t)=\sum_q \rho_{A,Q}^q(t) \otimes \bar P_B^{Q-q}$, where $\rho_{A,Q}^q$ is the deviation from MMS in region A associated with charge $q$ and also explicitly depends on the total charge $Q$. 
Therefore, the off-diagonal quadratic correction $\mathcal N_{B,\text{off}}^{(2)}$ can be derived as
\begin{equation}
	\begin{split}
		\mathcal N_{B,\text{off}}^{(2)}
		={}&\frac{1}{L_B(L_B-1)}\sum_{\substack{i,j\in B\\i\neq j}}\sum_Qw_Q^2
		\mathrm{Tr}\!\left[
		\bar\delta_Qc_i^\dagger c_j\bar\delta_Qc_j^\dagger c_i
		\right]\\
		={}&\sum_Qw_Q^2\sum_q\mathrm{Tr}\!\left[(\rho_{A,Q}^q)^2\right]
		\frac{1}{L_B(L_B-1)}\sum_{\substack{i,j\in B\\i\neq j}}
		\mathrm{Tr}\!\left[
		\bar P_B^{Q-q}c_i^\dagger c_j
		\bar P_B^{Q-q}c_j^\dagger c_i
		\right]\\
		={}&\sum_Qw_Q^2\sum_q\mathrm{Tr}\!\left[(\rho_{A,Q}^q)^2\right]
		\frac{1}{L_B(L_B-1)}
		\mathrm{Tr}\!\left[
		(\bar P_B^{Q-q})^2\hat Q_B(L_B-\hat Q_B)
		\right]\\
		={}&\sum_Qw_Q^2\sum_q\mathrm{Tr}\!\left[(\rho_{A,Q}^q)^2\right]
		\mathrm{Tr}\!\left[(\bar P_B^{Q-q})^2\right]
		\frac{(Q-q)(L_B-Q+q)}{L_B(L_B-1)}\\
		\simeq{}&\nu(1-\nu)\sum_Qw_Q^2
		\mathrm{Tr}\!\left[(\bar\delta_Q)^2\right].
	\end{split}
\end{equation}
In the last equality, we used $\frac{(Q-q)(L_B-Q+q)}{L_B(L_B-1)} = \nu(1-\nu)+\mathcal{O}(L_A/L_B) \simeq \nu(1-\nu)$, where $\nu=Q/(L_A+L_B)$ is the filling and $L_A/L_B \to 0$ for any finite time evolution in the thermodynamic limit. 

Following the similar approximation, the diagonal quadratic correction $\mathcal N_{B,\text{diag}}^{(2)}$ can be further derived as 
\begin{equation}
	\begin{aligned}
		\mathcal N_{B,\text{diag}}^{(2)}
		={}&\frac{1}{L_B}\sum_{i\in B}\sum_Qw_Qw_{Q+1}
		\mathrm{Tr}\!\left[
		\bar\delta_{Q+1}c_i^\dagger\bar\delta_Qc_i
		\right]\\
		={}&\sum_Qw_Qw_{Q+1}\sum_q
		\mathrm{Tr}\!\left[\rho_{A,Q+1}^q\rho_{A,Q}^q\right]
		\frac{1}{L_B}\sum_{i\in B}
		\mathrm{Tr}\!\left[
		\bar P_B^{Q+1-q}c_i^\dagger\bar P_B^{Q-q}c_i
		\right]\\
		={}&\sum_Qw_Qw_{Q+1}\sum_q
		\mathrm{Tr}\!\left[\rho_{A,Q+1}^q\rho_{A,Q}^q\right]
		\frac{1}{L_B}
		\frac{
			\sqrt{
				\mathrm{Tr}[P_B^{Q+1-q}\hat Q_B]\,
				\mathrm{Tr}[(L_B-\hat Q_B)P_B^{Q-q}]
			}
		}{
			\mathrm{Tr}[P_B^{Q+1-q}]
			\mathrm{Tr}[P_B^{Q-q}]
		}\\
		\simeq{}&\sqrt{\nu(1-\nu)}
		\sum_Qw_Qw_{Q+1}\sum_q
		\mathrm{Tr}\!\left[
		\frac{\rho_{A,Q+1}^q}{\sqrt{\mathrm{Tr}[P_B^{Q+1-q}]}}
		\frac{\rho_{A,Q}^q}{\sqrt{\mathrm{Tr}[P_B^{Q-q}]}}
		\right]\\
		\leq{}&\sqrt{\nu(1-\nu)}
		\sum_Qw_Qw_{Q+1}
		\sqrt{
			\sum_q\frac{\mathrm{Tr}[(\rho_{A,Q+1}^q)^2]}
			{\mathrm{Tr}[P_B^{Q+1-q}]}
		}
		\sqrt{
			\sum_q\frac{\mathrm{Tr}[(\rho_{A,Q}^q)^2]}
			{\mathrm{Tr}[P_B^{Q-q}]}
		}\\
		\leq{}&\sqrt{\nu(1-\nu)}
		\sum_Qw_Qw_{Q+1}
		\sqrt{\mathrm{Tr}[(\bar\delta_{Q+1})^2]}
		\sqrt{\mathrm{Tr}[(\bar\delta_Q)^2]}.
	\end{aligned}
\end{equation}

For notational convenience, we define $v_Q=w_Q\sqrt{\mathrm{Tr}[(\bar\delta_Q)^2]}$. 
The above results can then be recast as 
\begin{equation}
	\begin{gathered}
		\mathcal N_{B,\text{off}}^{(2)}\simeq\nu(1-\nu)\sum_Qv_Q^2,\\[2pt]
		\mathcal N_{B,\text{diag}}^{(2)}\leq\sqrt{\nu(1-\nu)}\sum_Qv_{Q+1}v_Q,\\[2pt]
		\mathcal Z_2^{(2)}=\sum_Qv_Q^2.
	\end{gathered}
\end{equation}

Substituting the above results into the inequality Eq.~\eqref{eq:R2-diff-inequality}, we have 
\begin{equation}
	\begin{aligned}
		\widetilde{\mathcal R}_2-\widetilde{\mathcal R}_{2,\text{aux}}
		&\geq \frac{\epsilon_\delta^2}{1+\epsilon_\delta^2}
		\left[\frac{\mathcal N_{B,\text{off}}^{(2)}}{\mathcal Z_2^{(2)}}
		-\left(\frac{\mathcal N_{B,\text{diag}}^{(2)}}{\mathcal Z_2^{(2)}}\right)^2
		-\widetilde{\mathcal R}_{2,\text{aux}}\right]\\
		&\geq \frac{\epsilon_\delta^2}{1+\epsilon_\delta^2}\nu(1-\nu)
		\left[\left(1-\left(\frac{\sum_Qv_{Q+1}v_Q}{\sum_Qv_Q^2}\right)^2\right)
		-\left(1-\left(\frac{\sum_Qu_{Q+1}u_Q}{\sum_Qu_Q^2}\right)^2\right)\right]\\
		&\geq -\frac{\epsilon_\delta^2}{1+\epsilon_\delta^2}\nu(1-\nu)
		\left[1-\left(\frac{\sum_Qu_{Q+1}u_Q}{\sum_Qu_Q^2}\right)^2\right]
		\simeq-\frac{\epsilon_\delta^2}{1+\epsilon_\delta^2}\widetilde{\mathcal R}_{2,\text{aux}},\\
		\Rightarrow \qquad
		&\widetilde{\mathcal R}_2(t)
		\gtrsim \frac{1}{1+\epsilon_\delta^2(t)}\widetilde{\mathcal R}_{2,\text{aux}}(t).
	\end{aligned}
	\label{eq:R2-conditional-lower-bound}
\end{equation}
Here $u_Q=w_Q/\sqrt{D_Q}$ denotes the purity-modified charge-sector weight, for which the auxiliary-state connected R\'enyi-2 correlation can be reexpressed as 
\begin{equation*}
	\widetilde{\mathcal R}_{2,\text{aux}}\simeq\nu(1-\nu)\left[1-\left(\sum_Qu_{Q+1}u_Q/\sum_Qu_Q^2\right)^2\right].
\end{equation*}
From the second line to the third line in Eq.~\eqref{eq:R2-conditional-lower-bound}, we used $0 \leq \left(\sum_Qv_{Q+1}v_Q/\sum_Qv_Q^2\right)^2 \leq 1$, which can be interpreted as the squared inner product of two normalized vectors. 
More precisely, we define $\mathbf v_1=(v_0,v_1,\ldots,v_L,0)^T$ and $\mathbf v_2=(0,v_0,v_1,\ldots,v_L)^T$. Since these vectors have equal norms, the Cauchy--Schwarz inequality gives $\left(\sum_Qv_{Q+1}v_Q/\sum_Qv_Q^2\right)^2=\left(\mathbf v_1\cdot\mathbf v_2/\|\mathbf v_1\|\|\mathbf v_2\|\right)^2\leq1$.

Since the time evolution of the charge-sector weight $w_Q$ follows Eq.~\eqref{eq:charge-sector-TE}, we can assume that $w_Q$ has a Gaussian distribution with the width $\sigma_Q$. 
In the regime $1 \ll \sigma_Q^2\ll L$, the width $\sigma_{Q,u}$ of the reweighted amplitude $u_Q$ satisfies $\sigma_{Q,u}\simeq\sigma_Q$, and the auxiliary-state correlation scales as $\widetilde{\mathcal R}_{2,\text{aux}} \sim \nu(1-\nu)/(2\sigma_Q^2)$. 
Since the mixed state finally relaxes to a well-defined stationary state under $\mathcal{L}_0 + \mathcal{V}$, $\epsilon_\delta^2(t)\to\epsilon^2<\infty$ reaches a stationary value, thus $1/[1+\epsilon_\delta^2(t)]$ contributes only a time-independent prefactor and does not affect the algebraic exponent. 
We finally obtain
\begin{equation}\label{eq:R2-algebraic-lower-bound}
	\widetilde{\mathcal R}_2(t)\gtrsim\frac{1}{1+\epsilon^2}\widetilde{\mathcal R}_{2,\text{aux}}(t)\gtrsim\frac{\nu(1-\nu)}{2(1+\epsilon^2)\sigma_Q^2(t)}.
\end{equation}
So far, we have proved that the auxiliary mixed state provides an algebraic lower bound for the exact connected R\'enyi-2 long-range correlation, up to a finite prefactor.

\subsection{Extension of the Lower Bound to the Connected R\'enyi-1 Correlator}
The above derivation of the lower bound for R\'enyi-2 correlations can be reproduced under the following two assumptions
\begin{equation}\label{eq:k1-two-assumptions}
	\begin{cases}
		\text{(i)} & \rho(0)=\bar P_{Q_0},\\
		\text{(ii)} & \displaystyle \rho_Q(t) =\bar P_Q + 
		\sum\nolimits_q \rho_{A,Q}^q(t) \otimes \bar P_B^{Q-q}.
	\end{cases}
\end{equation}
For (i), to examine how the SW-SSB order decays under local symmetry-breaking dynamics, we can always choose the initial state to be the MMS with a fixed symmetry charge, which respects the strong $U(1)$ symmetry and exhibits long-range SW-SSB order. 
For (ii), starting from the MMS initial state, the strong-symmetry-breaking channel acts locally on the mixed state, and the resulting disturbance spreads over the lattice under the Liouvillian dynamics $\mathcal{L}_0$. 
In the thermodynamic limit, for any finite evolution time, the disturbance remains confined within a finite subregion $A$ (Fig.~\ref{subfig:1}). 
Therefore, the time-evolved charge-sector state can be written in the form given in (ii). Note that we make no additional assumptions about the form of $\rho_{A,Q}^q(t)$. 

We now show that both assumptions are inherited by $\widetilde\rho\equiv\sqrt\rho$, allowing the same proof to be applied to the connected R\'enyi-1 correlator
\begin{equation}
	\begin{aligned}
		\widetilde{\mathcal R}_1
		&=\mathcal R_1^{\mathrm{off}}-(\mathcal R_1^{\mathrm{diag}})^2,
		\qquad \mathcal Z_1=\Tr[\rho]=1,\\
		\mathcal R_1^{\mathrm{off}}
		&=\frac{1}{L_B(L_B-1)}\sum_{\substack{i,j\in B\\i\neq j}}
		\frac{\Tr[\sqrt{\rho}\,c_i^\dagger c_j\sqrt{\rho}\,c_j^\dagger c_i]}{\mathcal Z_1},\\
		\mathcal R_1^{\mathrm{diag}}
		&=\frac{1}{L_B}\sum_{i\in B}
		\frac{\Tr[\sqrt{\rho}\,c_i^\dagger\sqrt{\rho}\,c_i]}{\mathcal Z_1}.
	\end{aligned}
	\label{eq:k1-correlator-definition}
\end{equation}
Indeed, the definitions differ from their R\'enyi-2 counterparts only by replacing $\rho$ with $\sqrt\rho$. Since the ratios are invariant under an overall rescaling of their state argument,
\begin{equation}\label{eq:k1-r1-r2-equivalence}
	\Tr[\widetilde\rho^2]=\Tr[\rho]=\mathcal Z_1,
	\qquad
	\widetilde{\mathcal R}_1(\rho) = \widetilde{\mathcal R}_2(\widetilde\rho).
\end{equation}

Assumption (i) in Eq.~\eqref{eq:k1-two-assumptions}: MMS initial condition. 
Since $P_{Q_0}$ is a projector, taking the square root changes the initial state only by an overall factor:
\begin{equation}
	\widetilde\rho(0)
	=\sqrt{\bar P_{Q_0}}
	=\frac{P_{Q_0}}{\sqrt{D_{Q_0}}}
	=\sqrt{D_{Q_0}}\,\bar P_{Q_0}
	\propto\bar P_{Q_0}.
	\label{eq:k1-mms-fixed-point}
\end{equation}
Thus, after an irrelevant normalization, assumption (i) also holds for $\widetilde\rho \equiv \sqrt{\rho}$.

Assumption (ii) in Eq.~\eqref{eq:k1-two-assumptions}: sector decomposition. 
Eq.~\eqref{EQ_MixedStateAnsatz} can be written as $\rho_Q=\sum_qR_{A,Q}^q\otimes\bar P_B^{Q-q}$, where $R_{A,Q}^q\equiv(D_{B,Q-q}/D_Q)P_A^q+\rho_{A,Q}^q\geq0$ is the complete charge-$q$ block in $A$ and $D_{B,Q-q}=\Tr_B[P_B^{Q-q}]$. Defining $\Pi_{Qq}\equiv P_A^q\otimes P_B^{Q-q}$, the orthogonality of different $q$ blocks gives
\begin{equation}
	\begin{aligned}
		\sqrt{\rho_Q}
		&=P_Q\sqrt{\rho_Q}P_Q,
		\qquad P_Q=\sum_q\Pi_{Qq},\\
		&=\sum_{q,q'}\Pi_{Qq}\sqrt{\rho_Q}\Pi_{Qq'}
		=\sum_q\sqrt{\Pi_{Qq}\rho_Q\Pi_{Qq}}\\
		&=\sum_q\sqrt{D_{B,Q-q}}\sqrt{R_{A,Q}^q}
		\otimes\bar P_B^{Q-q},\\
		\widetilde\rho_Q^{(1)}
		&\equiv\frac{\sqrt{\rho_Q}}{\Tr\sqrt{\rho_Q}}
		=\bar P_Q+\sum_q\widetilde\rho_{A,Q}^{q,(1)}
		\otimes\bar P_B^{Q-q}.
	\end{aligned}
	\label{eq:k1-square-root-spatial-decomposition}
\end{equation}
Here we used $\sqrt{\bar P_B^{Q-q}}=\sqrt{D_{B,Q-q}}\,\bar P_B^{Q-q}$. The operator $\widetilde\rho_{A,Q}^{q,(1)}$ denotes the deviation of the normalized positive charge-$q$ block from its MMS contribution and satisfies $\sum_q\Tr_A[\widetilde\rho_{A,Q}^{q,(1)}]=0$. Hence assumption (ii) is also preserved under $\rho\mapsto\sqrt\rho$.

Therefore, the connected R\'enyi-1 correlation of $\rho$ takes the same form as the connected R\'enyi-2 correlation of $\widetilde{\rho}=\sqrt{\rho}$. 
Since the total-charge sectors are mutually orthogonal, $\widetilde\rho=\sum_Qs_Q^{(1)}\widetilde\rho_Q^{(1)}$, and the R\'enyi-2 proof carries over under the replacements
\begin{equation}
	\rho\rightarrow\widetilde\rho\equiv\sqrt\rho,
	\qquad w_Q\rightarrow s_Q^{(1)}
	\equiv\sqrt{w_Q}\Tr\sqrt{\rho_Q},
	\qquad \rho_Q\rightarrow\widetilde\rho_Q^{(1)},
	\qquad \rho_{A,Q}^q\rightarrow\widetilde\rho_{A,Q}^{q,(1)}.
	\label{eq:k1-replacement-rule}
\end{equation}
The four entries respectively replace the full state, the charge-sector weights, the normalized charge-sector states, and the MMS deviation in region $A$. 
The corresponding auxiliary operator is $\widetilde\rho_{\mathrm{aux}}^{(1)}=\sum_Qs_Q^{(1)}\bar P_Q$, whose charge amplitude is $u_Q^{(1)}=s_Q^{(1)}/\sqrt{D_Q}=\sqrt{w_Q}\,\mathcal F(\rho_Q,\bar P_Q)$, with $\mathcal F(\rho_Q,\bar P_Q)=\Tr\sqrt{\rho_Q}/\sqrt{D_Q}$. In the sector-equilibrated regime, this fidelity is close to unity and varies weakly across the occupied charge window. Applying Eq.~\eqref{eq:R2-conditional-lower-bound} then gives
\begin{equation}\label{eq:k1-fidelity-lower-bound}
	\begin{aligned}
	\widetilde{\mathcal R}_1(\rho)
	&\gtrsim\overline{\mathcal F^2}\,
	\widetilde{\mathcal R}_2(\widetilde\rho_{\mathrm{aux}}^{(1)})\\
	&\simeq\overline{\mathcal F^2}\,
	\widetilde{\mathcal R}_1(\rho_{\mathrm{aux}})
	\simeq\overline{\mathcal F^2}\,
	\frac{\nu(1-\nu)}{4\sigma_Q^2}.
	\end{aligned}
\end{equation}
Here $\overline{\mathcal F^2}\equiv\sum_Qw_Q\mathcal F^2(\rho_Q,\bar P_Q)$ is a finite prefactor, and the last estimate assumes a broad Gaussian $w_Q$ with $1\ll\sigma_Q^2\ll L$. Thus the connected R\'enyi-1 correlator has the same algebraic lower envelope, $\widetilde{\mathcal R}_1\gtrsim\sigma_Q^{-2}$.

\begin{figure}[b]
	\begin{centering}
	\includegraphics[width=1\linewidth]{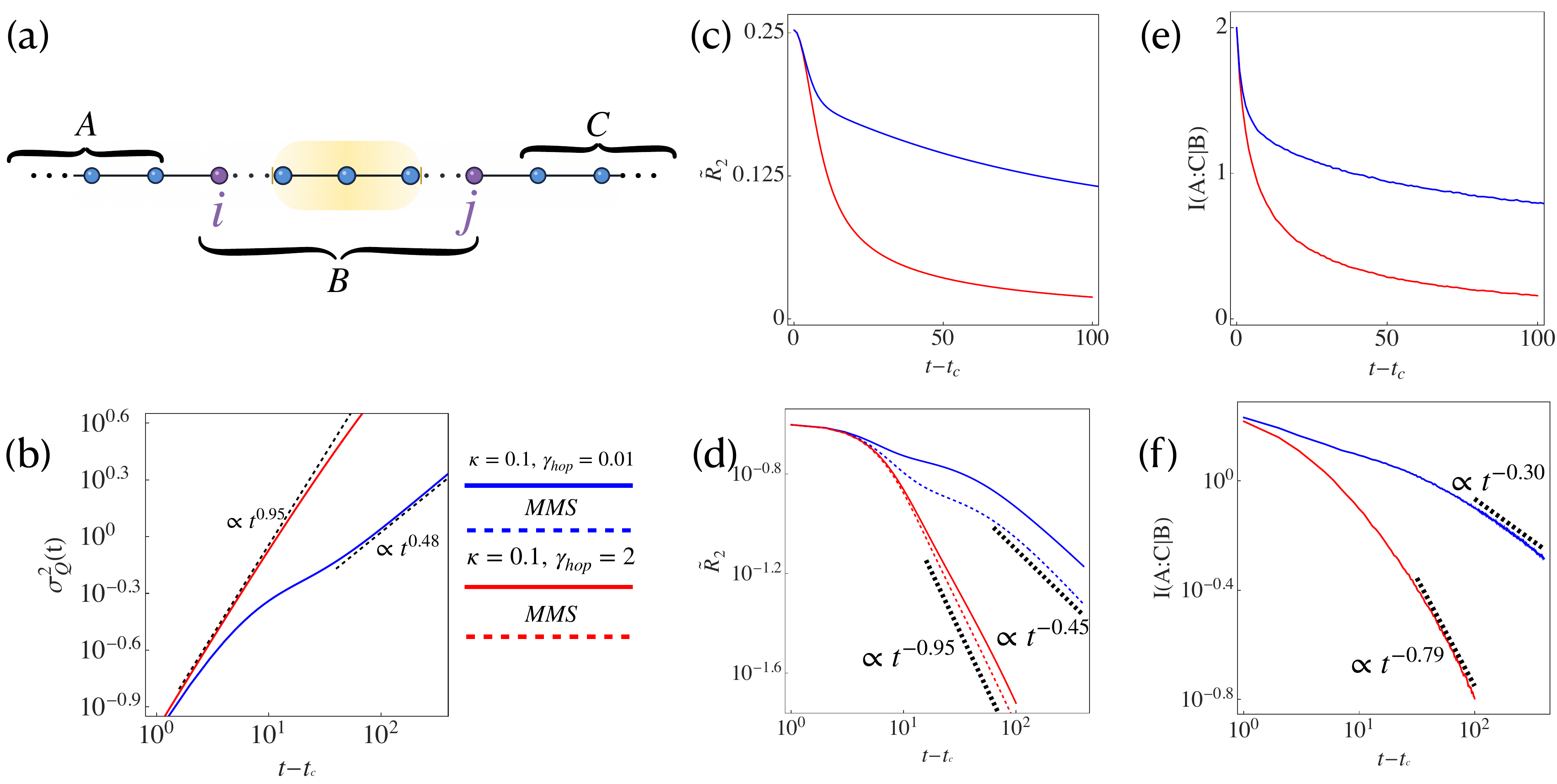}
	\par\end{centering}
	\protect\caption{\label{subfig:2}
	Numerical results for a spinless-fermion chain in Eq.~({\clr 15}) with $L=100$ in the main text, initialized in $\bar P_{L/2}$. A balanced leakage channel with $\kappa_+=\kappa_-=\kappa=0.1$ acts at $x=L/2$, with $\gamma_\phi=5$. Blue and red denote $\gamma_{\mathrm{hop}}=0.01$ and $2$, respectively. 
	{(a) Tripartition used for the conditional mutual information, with $A=\{1,\ldots,i\}$, $B=\{i+1,\ldots,j-1\}$, and $C=\{j,\ldots,L\}$.}
	(b) Charge variance on logarithmic scales; black dashed lines indicate $\sigma_Q^2\propto(t-t_c)^{0.48}$ (blue) and $(t-t_c)^{0.95}$ (red). 
	(c) Connected R\'enyi-2 correlator $\widetilde R_2(\rho,c_i^\dagger c_j)$ for $i=L/4$ and $j=3L/4$. Solid curves are numerical results, while colored dashed curves are the corresponding MMS predictions computed from the same $w_Q(t)$. 
	(d) Log-log plot of the same correlators; black dashed guides indicate $\widetilde R_2^{\rm aux}\propto(t-t_c)^{-0.45}$ (blue) and $(t-t_c)^{-0.95}$ (red). 
	{(e) Conditional mutual information $I(A:C\mid B)$ for the tripartition in panel (a).}
	{(f) Log-log plot of the same $I(A:C\mid B)$; black dashed lines indicate $(t-t_c)^{-0.30}$ (blue) and $(t-t_c)^{-0.79}$ (red).} }
\end{figure}

\subsection{Extension of the Lower Bound to General Connected R\'enyi-$k$ Correlators}
The $k=1$ and $k=2$ results show that the general connected R\'enyi-$k$ problem can be viewed as a R\'enyi-2 problem under the two assumptions in Eq.~\eqref{eq:k1-mms-fixed-point}. 
For each fixed $k\geq1$, define the positive operator $\widetilde\rho\equiv\rho^{k/2}$. Then
\begin{equation}
	\begin{aligned}
		\widetilde{\mathcal R}_k
		&=\mathcal R_k^{\mathrm{off}}-(\mathcal R_k^{\mathrm{diag}})^2,
		\qquad \mathcal Z_k=\Tr[\rho^k]=\Tr[\widetilde\rho^2],\\
		\mathcal R_k^{\mathrm{off}}
		&=\frac{1}{L_B(L_B-1)}\sum_{\substack{i,j\in B\\i\neq j}}
		\frac{\Tr[\widetilde\rho c_i^\dagger c_j\widetilde\rho c_j^\dagger c_i]}{\mathcal Z_k},\\
		\mathcal R_k^{\mathrm{diag}}
		&=\frac{1}{L_B}\sum_{i\in B}
		\frac{\Tr[\widetilde\rho c_i^\dagger\widetilde\rho c_i]}{\mathcal Z_k},
		\qquad
		\widetilde{\mathcal R}_k(\rho)
		=\widetilde{\mathcal R}_2(\widetilde\rho).
	\end{aligned}
\end{equation}
By the same argument used for $k=1$, the fixed-charge MMS condition and the conditional sector decomposition are preserved under $\rho\mapsto\rho^{k/2}$. The R\'enyi-2 result therefore gives directly
\begin{equation}\label{eq:renyi-k-conditional-lower-bound}
	\widetilde{\mathcal R}_k(\rho)
	\gtrsim\frac{1}{1+\epsilon_k^2(t)}
	\widetilde{\mathcal R}_{k,\mathrm{aux}}
	\simeq\frac{\nu(1-\nu)}
	{2[1+\epsilon_k^2(t)]\sigma_{Q,u^{(k)}}^2}.
\end{equation}
The quantities entering this bound are obtained from their R\'enyi-2 counterparts by
\begin{equation}\label{eq:renyi-k-sector-amplitude}
	\begin{aligned}
	\widetilde\rho
	&=\sum_Qw_Q^{k/2}\rho_Q^{k/2}
	=\sum_Qs_Q^{(k)}\widetilde\rho_Q^{(k)},
	& s_Q^{(k)}&\equiv w_Q^{k/2}\Tr[\rho_Q^{k/2}],\\
	\widetilde\rho_Q^{(k)}
	&\equiv\frac{\rho_Q^{k/2}}{\Tr[\rho_Q^{k/2}]},
	& \widetilde\rho_{\mathrm{aux}}&\equiv\sum_Qs_Q^{(k)}\bar P_Q,\\
	u_Q^{(k)}
	&\equiv\frac{s_Q^{(k)}}{\sqrt{D_Q}}
	=\frac{w_Q^{k/2}\Tr[\rho_Q^{k/2}]}{\sqrt{D_Q}},
	& \widetilde{\mathcal R}_{k,\mathrm{aux}}
	&\equiv\widetilde{\mathcal R}_2(\widetilde\rho_{\mathrm{aux}}),\\
	\mathcal Z_k^{(0)}
	&\equiv\sum_Q[u_Q^{(k)}]^2,
	& \epsilon_k^2(t)&\equiv
	\frac{\mathcal Z_k^{(2)}}{\mathcal Z_k^{(0)}}.
	\end{aligned}
\end{equation}
Here $\mathcal Z_k^{(2)}\equiv\Tr[(\widetilde\rho-\widetilde\rho_{\mathrm{aux}})^2]$, and sectorwise tracelessness gives $\mathcal Z_k=\mathcal Z_k^{(0)}+\mathcal Z_k^{(2)}$. The cases $k=1$ and $k=2$ reproduce $u_Q^{(1)}=\sqrt{w_Q}\mathcal F(\rho_Q,\bar P_Q)$ and $u_Q^{(2)}=w_Q/\sqrt{D_Q}$, respectively. The auxiliary neighboring-sector overlap is
\begin{equation}\label{eq:renyi-k-auxiliary-overlap}
	\widetilde{\mathcal R}_{k,\mathrm{aux}}
	\simeq\nu(1-\nu)
	\left[1-\left(
	\frac{\sum_Qu_Q^{(k)}u_{Q+1}^{(k)}}
	{\sum_Q[u_Q^{(k)}]^2}\right)^2\right]
	\simeq\frac{\nu(1-\nu)}{2\sigma_{Q,u^{(k)}}^2}.
\end{equation}
Within the auxiliary-state approximation $\rho_Q\simeq\bar P_Q$, one has $u_Q^{(k)}\simeq w_Q^{k/2}D_Q^{(1-k)/2}$. For a broad Gaussian $w_Q$ with $\sigma_Q^2\ll L$, the corresponding width is
\begin{equation}
	\sigma_{Q,u^{(k)}}^2
	=\frac{2\sigma_Q^2}{k}\left[1+O\left(\frac{\sigma_Q^2}{L}\right)\right]
	\simeq\frac{2\sigma_Q^2}{k}.
\end{equation}
Once sector equilibration is reached within the pre-saturation regime, $\epsilon_k^2(t)$ contributes only a finite prefactor. Thus, for every fixed $k$, charge-space broadening yields the same algebraic lower envelope, $\widetilde{\mathcal R}_k\gtrsim\sigma_Q^{-2}$, while $k$ changes only its leading prefactor. 

\section{Numerical data for the R\'enyi-2 correlation and the conditional mutual information}

Thus far, we have shown that the charge-space broadening controls the algebraic lower envelope of connected R\'enyi-$k$ correlators for any positive integer $k$. 
In the main text, we verified this relation numerically using the connected R\'enyi-1 correlator. 
For completeness, Figs.~\ref{subfig:2}(b)-(d) present the corresponding R\'enyi-2 results for the same strong-symmetry-breaking dynamics. 
The solid curves show the numerical results for connected long-range R\'enyi-2 correlation, for comparison, the colored dashed curves are the corresponding correlation obtained from the auxiliary mixed state $\rho_{\text{aux}}(t)$ given by Eq.~\eqref{EQ_AuxMixState}. 
It shows that the auxiliary-state long-range correlation decay algebraically as $\widetilde R_2^{\rm aux}\propto(t-t_c)^{-0.45}$ for $\gamma_{\mathrm{hop}}=0.01$ and $\widetilde R_2^{\rm aux}\propto(t-t_c)^{-0.95}$ for $\gamma_{\mathrm{hop}}=2$. 
In both cases, the actual R\'enyi-2 long-range correlation remains lower-bounded by the auxiliary-state correlation and decays more slowly than its auxiliary-state counterpart.
This confirms that the auxiliary-state result provides an algebraically decaying lower bound for the actual long-range correlation. 

We further examine the conditional mutual information (CMI) defined with the tripartition illustrated in Fig.~\ref{subfig:2}(a). 
For the reduced density matrix $\rho_X$ of a subregion $X$, we define the von Neumann entropy as $S(X)=-\mathrm{Tr}[\rho_X\ln\rho_X]$. 
The conditional mutual information is then
\begin{equation}
	I(A:C\mid B)=S(AB)+S(BC)-S(B)-S(ABC).
	\label{eq:CMI-definition}
\end{equation}
The numerical results shown in Figs.~\ref{subfig:2}(e) and (f) exhibit algebraic relaxation, with $I(A:C\mid B)\propto(t-t_c)^{-0.30}$ for $\gamma_{\mathrm{hop}}=0.01$ and $I(A:C\mid B)\propto(t-t_c)^{-0.79}$ for $\gamma_{\mathrm{hop}}=2$. 
Both exponents are smaller than the corresponding auxiliary-state R\'enyi-2 decay exponents, $0.45$ and $0.95$, respectively, as shown in Fig.~\ref{subfig:2}(d). 
It is therefore evident that the CMI decays more slowly than the R''enyi-1 and R''enyi-2 correlations. 
We emphasize, however, that this slower decay is a numerical observation rather than a consequence of the bound derived above. A full analytical theory of the CMI dynamics is left for future work.


\begin{thebibliography}{57}%
\makeatletter
\providecommand \@ifxundefined [1]{%
 \@ifx{#1\undefined}
}%
\providecommand \@ifnum [1]{%
 \ifnum #1\expandafter \@firstoftwo
 \else \expandafter \@secondoftwo
 \fi
}%
\providecommand \@ifx [1]{%
 \ifx #1\expandafter \@firstoftwo
 \else \expandafter \@secondoftwo
 \fi
}%
\providecommand \natexlab [1]{#1}%
\providecommand \enquote  [1]{``#1''}%
\providecommand \bibnamefont  [1]{#1}%
\providecommand \bibfnamefont [1]{#1}%
\providecommand \citenamefont [1]{#1}%
\providecommand \href@noop [0]{\@secondoftwo}%
\providecommand \href [0]{\begingroup \@sanitize@url \@href}%
\providecommand \@href[1]{\@@startlink{#1}\@@href}%
\providecommand \@@href[1]{\endgroup#1\@@endlink}%
\providecommand \@sanitize@url [0]{\catcode `\\12\catcode `\$12\catcode
  `\&12\catcode `\#12\catcode `\^12\catcode `\_12\catcode `\%12\relax}%
\providecommand \@@startlink[1]{}%
\providecommand \@@endlink[0]{}%
\providecommand \url  [0]{\begingroup\@sanitize@url \@url }%
\providecommand \@url [1]{\endgroup\@href {#1}{\urlprefix }}%
\providecommand \urlprefix  [0]{URL }%
\providecommand \Eprint [0]{\href }%
\providecommand \doibase [0]{https://doi.org/}%
\providecommand \selectlanguage [0]{\@gobble}%
\providecommand \bibinfo  [0]{\@secondoftwo}%
\providecommand \bibfield  [0]{\@secondoftwo}%
\providecommand \translation [1]{[#1]}%
\providecommand \BibitemOpen [0]{}%
\providecommand \bibitemStop [0]{}%
\providecommand \bibitemNoStop [0]{.\EOS\space}%
\providecommand \EOS [0]{\spacefactor3000\relax}%
\providecommand \BibitemShut  [1]{\csname bibitem#1\endcsname}%
\let\auto@bib@innerbib\@empty
\bibitem [{\citenamefont {Buča}\ and\ \citenamefont
  {Prosen}(2012)}]{Prosen2012IOP}%
  \BibitemOpen
  \bibfield  {author} {\bibinfo {author} {\bibfnamefont {B.}~\bibnamefont
  {Buča}}\ and\ \bibinfo {author} {\bibfnamefont {T.}~\bibnamefont {Prosen}},\
  }\bibfield  {title} {\bibinfo {title} {A note on symmetry reductions of the
  lindblad equation: transport in constrained open spin chains},\ }\href
  {https://doi.org/10.1088/1367-2630/14/7/073007} {\bibfield  {journal}
  {\bibinfo  {journal} {New Journal of Physics}\ }\textbf {\bibinfo {volume}
  {14}},\ \bibinfo {pages} {073007} (\bibinfo {year} {2012})}\BibitemShut
  {NoStop}%
\bibitem [{\citenamefont {Albert}\ and\ \citenamefont
  {Jiang}(2014)}]{JiangL2014PRA}%
  \BibitemOpen
  \bibfield  {author} {\bibinfo {author} {\bibfnamefont {V.~V.}\ \bibnamefont
  {Albert}}\ and\ \bibinfo {author} {\bibfnamefont {L.}~\bibnamefont {Jiang}},\
  }\bibfield  {title} {\bibinfo {title} {Symmetries and conserved quantities in
  lindblad master equations},\ }\href
  {https://doi.org/10.1103/PhysRevA.89.022118} {\bibfield  {journal} {\bibinfo
  {journal} {Phys. Rev. A}\ }\textbf {\bibinfo {volume} {89}},\ \bibinfo
  {pages} {022118} (\bibinfo {year} {2014})}\BibitemShut {NoStop}%
\bibitem [{\citenamefont {Coser}\ and\ \citenamefont
  {P{\'{e}}rez-Garc{\'{i}}a}(2019)}]{Coser2019Quantum}%
  \BibitemOpen
  \bibfield  {author} {\bibinfo {author} {\bibfnamefont {A.}~\bibnamefont
  {Coser}}\ and\ \bibinfo {author} {\bibfnamefont {D.}~\bibnamefont
  {P{\'{e}}rez-Garc{\'{i}}a}},\ }\bibfield  {title} {\bibinfo {title}
  {Classification of phases for mixed states via fast dissipative evolution},\
  }\href {https://doi.org/10.22331/q-2019-08-12-174} {\bibfield  {journal}
  {\bibinfo  {journal} {{Quantum}}\ }\textbf {\bibinfo {volume} {3}},\ \bibinfo
  {pages} {174} (\bibinfo {year} {2019})}\BibitemShut {NoStop}%
\bibitem [{\citenamefont {Lieu}\ \emph {et~al.}(2020)\citenamefont {Lieu},
  \citenamefont {Belyansky}, \citenamefont {Young}, \citenamefont {Lundgren},
  \citenamefont {Albert},\ and\ \citenamefont {Gorshkov}}]{Gorshkov2020PRL}%
  \BibitemOpen
  \bibfield  {author} {\bibinfo {author} {\bibfnamefont {S.}~\bibnamefont
  {Lieu}}, \bibinfo {author} {\bibfnamefont {R.}~\bibnamefont {Belyansky}},
  \bibinfo {author} {\bibfnamefont {J.~T.}\ \bibnamefont {Young}}, \bibinfo
  {author} {\bibfnamefont {R.}~\bibnamefont {Lundgren}}, \bibinfo {author}
  {\bibfnamefont {V.~V.}\ \bibnamefont {Albert}},\ and\ \bibinfo {author}
  {\bibfnamefont {A.~V.}\ \bibnamefont {Gorshkov}},\ }\bibfield  {title}
  {\bibinfo {title} {Symmetry breaking and error correction in open quantum
  systems},\ }\href {https://doi.org/10.1103/PhysRevLett.125.240405} {\bibfield
   {journal} {\bibinfo  {journal} {Phys. Rev. Lett.}\ }\textbf {\bibinfo
  {volume} {125}},\ \bibinfo {pages} {240405} (\bibinfo {year}
  {2020})}\BibitemShut {NoStop}%
\bibitem [{\citenamefont {de~Groot}\ \emph {et~al.}(2022)\citenamefont
  {de~Groot}, \citenamefont {Turzillo},\ and\ \citenamefont
  {Schuch}}]{Schuch2022Quantum}%
  \BibitemOpen
  \bibfield  {author} {\bibinfo {author} {\bibfnamefont {C.}~\bibnamefont
  {de~Groot}}, \bibinfo {author} {\bibfnamefont {A.}~\bibnamefont {Turzillo}},\
  and\ \bibinfo {author} {\bibfnamefont {N.}~\bibnamefont {Schuch}},\
  }\bibfield  {title} {\bibinfo {title} {Symmetry {P}rotected {T}opological
  {O}rder in {O}pen {Q}uantum {S}ystems},\ }\href
  {https://doi.org/10.22331/q-2022-11-10-856} {\bibfield  {journal} {\bibinfo
  {journal} {{Quantum}}\ }\textbf {\bibinfo {volume} {6}},\ \bibinfo {pages}
  {856} (\bibinfo {year} {2022})}\BibitemShut {NoStop}%
\bibitem [{\citenamefont {Lee}\ \emph {et~al.}(2023)\citenamefont {Lee},
  \citenamefont {Jian},\ and\ \citenamefont {Xu}}]{XuCK2023PRXQ}%
  \BibitemOpen
  \bibfield  {author} {\bibinfo {author} {\bibfnamefont {J.~Y.}\ \bibnamefont
  {Lee}}, \bibinfo {author} {\bibfnamefont {C.-M.}\ \bibnamefont {Jian}},\ and\
  \bibinfo {author} {\bibfnamefont {C.}~\bibnamefont {Xu}},\ }\bibfield
  {title} {\bibinfo {title} {Quantum criticality under decoherence or weak
  measurement},\ }\href {https://doi.org/10.1103/PRXQuantum.4.030317}
  {\bibfield  {journal} {\bibinfo  {journal} {PRX Quantum}\ }\textbf {\bibinfo
  {volume} {4}},\ \bibinfo {pages} {030317} (\bibinfo {year}
  {2023})}\BibitemShut {NoStop}%
\bibitem [{\citenamefont {Ma}\ \emph {et~al.}(2025)\citenamefont {Ma},
  \citenamefont {Zhang}, \citenamefont {Bi}, \citenamefont {Cheng},\ and\
  \citenamefont {Wang}}]{WangC2025PRX}%
  \BibitemOpen
  \bibfield  {author} {\bibinfo {author} {\bibfnamefont {R.}~\bibnamefont
  {Ma}}, \bibinfo {author} {\bibfnamefont {J.-H.}\ \bibnamefont {Zhang}},
  \bibinfo {author} {\bibfnamefont {Z.}~\bibnamefont {Bi}}, \bibinfo {author}
  {\bibfnamefont {M.}~\bibnamefont {Cheng}},\ and\ \bibinfo {author}
  {\bibfnamefont {C.}~\bibnamefont {Wang}},\ }\bibfield  {title} {\bibinfo
  {title} {Topological phases with average symmetries: The decohered, the
  disordered, and the intrinsic},\ }\href
  {https://doi.org/10.1103/PhysRevX.15.021062} {\bibfield  {journal} {\bibinfo
  {journal} {Phys. Rev. X}\ }\textbf {\bibinfo {volume} {15}},\ \bibinfo
  {pages} {021062} (\bibinfo {year} {2025})}\BibitemShut {NoStop}%
\bibitem [{\citenamefont {Lessa}\ \emph
  {et~al.}(2025{\natexlab{a}})\citenamefont {Lessa}, \citenamefont {Cheng},\
  and\ \citenamefont {Wang}}]{Lessa2025PRX}%
  \BibitemOpen
  \bibfield  {author} {\bibinfo {author} {\bibfnamefont {L.~A.}\ \bibnamefont
  {Lessa}}, \bibinfo {author} {\bibfnamefont {M.}~\bibnamefont {Cheng}},\ and\
  \bibinfo {author} {\bibfnamefont {C.}~\bibnamefont {Wang}},\ }\bibfield
  {title} {\bibinfo {title} {Mixed-state quantum anomaly and multipartite
  entanglement},\ }\href {https://doi.org/10.1103/PhysRevX.15.011069}
  {\bibfield  {journal} {\bibinfo  {journal} {Phys. Rev. X}\ }\textbf {\bibinfo
  {volume} {15}},\ \bibinfo {pages} {011069} (\bibinfo {year}
  {2025}{\natexlab{a}})}\BibitemShut {NoStop}%
\bibitem [{\citenamefont {Guo}\ \emph {et~al.}(2025{\natexlab{a}})\citenamefont
  {Guo}, \citenamefont {Zhang}, \citenamefont {Zhang}, \citenamefont {Yang},\
  and\ \citenamefont {Bi}}]{BiZhen2025PRX}%
  \BibitemOpen
  \bibfield  {author} {\bibinfo {author} {\bibfnamefont {Y.}~\bibnamefont
  {Guo}}, \bibinfo {author} {\bibfnamefont {J.-H.}\ \bibnamefont {Zhang}},
  \bibinfo {author} {\bibfnamefont {H.-R.}\ \bibnamefont {Zhang}}, \bibinfo
  {author} {\bibfnamefont {S.}~\bibnamefont {Yang}},\ and\ \bibinfo {author}
  {\bibfnamefont {Z.}~\bibnamefont {Bi}},\ }\bibfield  {title} {\bibinfo
  {title} {Locally purified density operators for symmetry-protected
  topological phases in mixed states},\ }\href
  {https://doi.org/10.1103/PhysRevX.15.021060} {\bibfield  {journal} {\bibinfo
  {journal} {Phys. Rev. X}\ }\textbf {\bibinfo {volume} {15}},\ \bibinfo
  {pages} {021060} (\bibinfo {year} {2025}{\natexlab{a}})}\BibitemShut
  {NoStop}%
\bibitem [{\citenamefont {Huang}\ \emph
  {et~al.}(2025{\natexlab{a}})\citenamefont {Huang}, \citenamefont {Diehl},\
  and\ \citenamefont {Sun}}]{XQSun2025PRL}%
  \BibitemOpen
  \bibfield  {author} {\bibinfo {author} {\bibfnamefont {Z.-M.}\ \bibnamefont
  {Huang}}, \bibinfo {author} {\bibfnamefont {S.}~\bibnamefont {Diehl}},\ and\
  \bibinfo {author} {\bibfnamefont {X.-Q.}\ \bibnamefont {Sun}},\ }\bibfield
  {title} {\bibinfo {title} {Topological response in open quantum systems with
  weak symmetries},\ }\href {https://doi.org/10.1103/qg2g-nz88} {\bibfield
  {journal} {\bibinfo  {journal} {Phys. Rev. Lett.}\ }\textbf {\bibinfo
  {volume} {135}},\ \bibinfo {pages} {213002} (\bibinfo {year}
  {2025}{\natexlab{a}})}\BibitemShut {NoStop}%
\bibitem [{\citenamefont {Kuno}\ \emph {et~al.}(2024)\citenamefont {Kuno},
  \citenamefont {Orito},\ and\ \citenamefont {Ichinose}}]{IKuo2024PRB}%
  \BibitemOpen
  \bibfield  {author} {\bibinfo {author} {\bibfnamefont {Y.}~\bibnamefont
  {Kuno}}, \bibinfo {author} {\bibfnamefont {T.}~\bibnamefont {Orito}},\ and\
  \bibinfo {author} {\bibfnamefont {I.}~\bibnamefont {Ichinose}},\ }\bibfield
  {title} {\bibinfo {title} {Strong-to-weak symmetry breaking states in
  stochastic dephasing stabilizer circuits},\ }\href
  {https://doi.org/10.1103/PhysRevB.110.094106} {\bibfield  {journal} {\bibinfo
   {journal} {Phys. Rev. B}\ }\textbf {\bibinfo {volume} {110}},\ \bibinfo
  {pages} {094106} (\bibinfo {year} {2024})}\BibitemShut {NoStop}%
\bibitem [{\citenamefont {Guo}\ \emph {et~al.}(2025{\natexlab{b}})\citenamefont
  {Guo}, \citenamefont {Yang},\ and\ \citenamefont
  {Yu}}]{XueJia2025PRXQuantum}%
  \BibitemOpen
  \bibfield  {author} {\bibinfo {author} {\bibfnamefont {Y.}~\bibnamefont
  {Guo}}, \bibinfo {author} {\bibfnamefont {S.}~\bibnamefont {Yang}},\ and\
  \bibinfo {author} {\bibfnamefont {X.-J.}\ \bibnamefont {Yu}},\ }\bibfield
  {title} {\bibinfo {title} {Quantum strong-to-weak spontaneous symmetry
  breaking in decohered one-dimensional critical states},\ }\href
  {https://doi.org/10.1103/4vs5-l54f} {\bibfield  {journal} {\bibinfo
  {journal} {PRX Quantum}\ }\textbf {\bibinfo {volume} {6}},\ \bibinfo {pages}
  {040311} (\bibinfo {year} {2025}{\natexlab{b}})}\BibitemShut {NoStop}%
\bibitem [{\citenamefont {Lessa}\ \emph
  {et~al.}(2025{\natexlab{b}})\citenamefont {Lessa}, \citenamefont {Ma},
  \citenamefont {Zhang}, \citenamefont {Bi}, \citenamefont {Cheng},\ and\
  \citenamefont {Wang}}]{Lessa2025PRXQ}%
  \BibitemOpen
  \bibfield  {author} {\bibinfo {author} {\bibfnamefont {L.~A.}\ \bibnamefont
  {Lessa}}, \bibinfo {author} {\bibfnamefont {R.}~\bibnamefont {Ma}}, \bibinfo
  {author} {\bibfnamefont {J.-H.}\ \bibnamefont {Zhang}}, \bibinfo {author}
  {\bibfnamefont {Z.}~\bibnamefont {Bi}}, \bibinfo {author} {\bibfnamefont
  {M.}~\bibnamefont {Cheng}},\ and\ \bibinfo {author} {\bibfnamefont
  {C.}~\bibnamefont {Wang}},\ }\bibfield  {title} {\bibinfo {title}
  {Strong-to-weak spontaneous symmetry breaking in mixed quantum states},\
  }\href {https://doi.org/10.1103/PRXQuantum.6.010344} {\bibfield  {journal}
  {\bibinfo  {journal} {PRX Quantum}\ }\textbf {\bibinfo {volume} {6}},\
  \bibinfo {pages} {010344} (\bibinfo {year} {2025}{\natexlab{b}})}\BibitemShut
  {NoStop}%
\bibitem [{\citenamefont {Weinstein}(2025)}]{Weinstein2025PRL}%
  \BibitemOpen
  \bibfield  {author} {\bibinfo {author} {\bibfnamefont {Z.}~\bibnamefont
  {Weinstein}},\ }\bibfield  {title} {\bibinfo {title} {Efficient detection of
  strong-to-weak spontaneous symmetry breaking via the r\'enyi-1 correlator},\
  }\href {https://doi.org/10.1103/PhysRevLett.134.150405} {\bibfield  {journal}
  {\bibinfo  {journal} {Phys. Rev. Lett.}\ }\textbf {\bibinfo {volume} {134}},\
  \bibinfo {pages} {150405} (\bibinfo {year} {2025})}\BibitemShut {NoStop}%
\bibitem [{\citenamefont {Liu}\ \emph {et~al.}(2025)\citenamefont {Liu},
  \citenamefont {Chen}, \citenamefont {Zhang}, \citenamefont {Zhou},\ and\
  \citenamefont {Zhang}}]{ZhangPF2025CP}%
  \BibitemOpen
  \bibfield  {author} {\bibinfo {author} {\bibfnamefont {Z.}~\bibnamefont
  {Liu}}, \bibinfo {author} {\bibfnamefont {L.}~\bibnamefont {Chen}}, \bibinfo
  {author} {\bibfnamefont {Y.}~\bibnamefont {Zhang}}, \bibinfo {author}
  {\bibfnamefont {S.}~\bibnamefont {Zhou}},\ and\ \bibinfo {author}
  {\bibfnamefont {P.}~\bibnamefont {Zhang}},\ }\bibfield  {title} {\bibinfo
  {title} {Diagnosing strong-to-weak symmetry breaking via wightman
  correlators},\ }\href {https://doi.org/10.1038/s42005-025-02199-7} {\bibfield
   {journal} {\bibinfo  {journal} {Communications Physics}\ }\textbf {\bibinfo
  {volume} {8}},\ \bibinfo {pages} {274} (\bibinfo {year} {2025})}\BibitemShut
  {NoStop}%
\bibitem [{\citenamefont {Sun}\ \emph {et~al.}(2025)\citenamefont {Sun},
  \citenamefont {Zhang},\ and\ \citenamefont {Feng}}]{FengL2025PRL}%
  \BibitemOpen
  \bibfield  {author} {\bibinfo {author} {\bibfnamefont {N.}~\bibnamefont
  {Sun}}, \bibinfo {author} {\bibfnamefont {P.}~\bibnamefont {Zhang}},\ and\
  \bibinfo {author} {\bibfnamefont {L.}~\bibnamefont {Feng}},\ }\bibfield
  {title} {\bibinfo {title} {Scheme to detect the strong-to-weak symmetry
  breaking via randomized measurements},\ }\href
  {https://doi.org/10.1103/7p5x-7yqb} {\bibfield  {journal} {\bibinfo
  {journal} {Phys. Rev. Lett.}\ }\textbf {\bibinfo {volume} {135}},\ \bibinfo
  {pages} {090403} (\bibinfo {year} {2025})}\BibitemShut {NoStop}%
\bibitem [{\citenamefont {Feng}\ \emph {et~al.}(2025)\citenamefont {Feng},
  \citenamefont {Cheng},\ and\ \citenamefont {Ippoliti}}]{Ippoliti2025PRL}%
  \BibitemOpen
  \bibfield  {author} {\bibinfo {author} {\bibfnamefont {X.}~\bibnamefont
  {Feng}}, \bibinfo {author} {\bibfnamefont {Z.}~\bibnamefont {Cheng}},\ and\
  \bibinfo {author} {\bibfnamefont {M.}~\bibnamefont {Ippoliti}},\ }\bibfield
  {title} {\bibinfo {title} {Hardness of observing strong-to-weak symmetry
  breaking},\ }\href {https://doi.org/10.1103/1xzd-g9s5} {\bibfield  {journal}
  {\bibinfo  {journal} {Phys. Rev. Lett.}\ }\textbf {\bibinfo {volume} {135}},\
  \bibinfo {pages} {200402} (\bibinfo {year} {2025})}\BibitemShut {NoStop}%
\bibitem [{\citenamefont {Sala}\ \emph {et~al.}(2024)\citenamefont {Sala},
  \citenamefont {Gopalakrishnan}, \citenamefont {Oshikawa},\ and\ \citenamefont
  {You}}]{YouYZ2024PRB}%
  \BibitemOpen
  \bibfield  {author} {\bibinfo {author} {\bibfnamefont {P.}~\bibnamefont
  {Sala}}, \bibinfo {author} {\bibfnamefont {S.}~\bibnamefont
  {Gopalakrishnan}}, \bibinfo {author} {\bibfnamefont {M.}~\bibnamefont
  {Oshikawa}},\ and\ \bibinfo {author} {\bibfnamefont {Y.}~\bibnamefont
  {You}},\ }\bibfield  {title} {\bibinfo {title} {Spontaneous strong symmetry
  breaking in open systems: Purification perspective},\ }\href
  {https://doi.org/10.1103/PhysRevB.110.155150} {\bibfield  {journal} {\bibinfo
   {journal} {Phys. Rev. B}\ }\textbf {\bibinfo {volume} {110}},\ \bibinfo
  {pages} {155150} (\bibinfo {year} {2024})}\BibitemShut {NoStop}%
\bibitem [{\citenamefont {Gu}\ \emph {et~al.}(2025)\citenamefont {Gu},
  \citenamefont {Wang},\ and\ \citenamefont {Wang}}]{WangZhong2025PRB}%
  \BibitemOpen
  \bibfield  {author} {\bibinfo {author} {\bibfnamefont {D.}~\bibnamefont
  {Gu}}, \bibinfo {author} {\bibfnamefont {Z.}~\bibnamefont {Wang}},\ and\
  \bibinfo {author} {\bibfnamefont {Z.}~\bibnamefont {Wang}},\ }\bibfield
  {title} {\bibinfo {title} {Spontaneous symmetry breaking in open quantum
  systems: Strong, weak, and strong-to-weak},\ }\href
  {https://doi.org/10.1103/3g6d-gn7b} {\bibfield  {journal} {\bibinfo
  {journal} {Phys. Rev. B}\ }\textbf {\bibinfo {volume} {112}},\ \bibinfo
  {pages} {245123} (\bibinfo {year} {2025})}\BibitemShut {NoStop}%
\bibitem [{\citenamefont {Guo}\ and\ \citenamefont
  {Yang}(2025)}]{ShuoYang2025PRB}%
  \BibitemOpen
  \bibfield  {author} {\bibinfo {author} {\bibfnamefont {Y.}~\bibnamefont
  {Guo}}\ and\ \bibinfo {author} {\bibfnamefont {S.}~\bibnamefont {Yang}},\
  }\bibfield  {title} {\bibinfo {title} {Strong-to-weak spontaneous symmetry
  breaking meets average symmetry-protected topological order},\ }\href
  {https://doi.org/10.1103/PhysRevB.111.L201108} {\bibfield  {journal}
  {\bibinfo  {journal} {Phys. Rev. B}\ }\textbf {\bibinfo {volume} {111}},\
  \bibinfo {pages} {L201108} (\bibinfo {year} {2025})}\BibitemShut {NoStop}%
\bibitem [{\citenamefont {Ziereis}\ \emph {et~al.}(2025)\citenamefont
  {Ziereis}, \citenamefont {Moudgalya},\ and\ \citenamefont
  {Knap}}]{Ziereis2025arXiv}%
  \BibitemOpen
  \bibfield  {author} {\bibinfo {author} {\bibfnamefont {N.}~\bibnamefont
  {Ziereis}}, \bibinfo {author} {\bibfnamefont {S.}~\bibnamefont {Moudgalya}},\
  and\ \bibinfo {author} {\bibfnamefont {M.}~\bibnamefont {Knap}},\ }\href
  {https://doi.org/10.48550/arXiv.2509.09669} {\bibinfo {title} {Strong-to-weak
  symmetry breaking phases in steady states of quantum operations}} (\bibinfo
  {year} {2025}),\ \Eprint {https://arxiv.org/abs/2509.09669} {arXiv:2509.09669
  [cond-mat.stat-mech]} \BibitemShut {NoStop}%
\bibitem [{\citenamefont {Wang}(2026)}]{WangChong2026Review}%
  \BibitemOpen
  \bibfield  {author} {\bibinfo {author} {\bibfnamefont {C.}~\bibnamefont
  {Wang}},\ }\href@noop {} {\bibinfo {title} {Strong-to-weak spontaneous
  symmetry breaking}} (\bibinfo {year} {2026}),\ \Eprint
  {https://arxiv.org/abs/arXiv:2606.02555} {arXiv:2606.02555} \BibitemShut
  {NoStop}%
\bibitem [{\citenamefont {Zhang}\ \emph {et~al.}(2024)\citenamefont {Zhang},
  \citenamefont {Xu},\ and\ \citenamefont {Xu}}]{JHZhang2024}%
  \BibitemOpen
  \bibfield  {author} {\bibinfo {author} {\bibfnamefont {J.-H.}\ \bibnamefont
  {Zhang}}, \bibinfo {author} {\bibfnamefont {C.}~\bibnamefont {Xu}},\ and\
  \bibinfo {author} {\bibfnamefont {Y.}~\bibnamefont {Xu}},\ }\href@noop {}
  {\bibinfo {title} {Fluctuation-dissipation theorem and information geometry
  in open quantum systems}} (\bibinfo {year} {2024}),\ \Eprint
  {https://arxiv.org/abs/arXiv:2409.18944} {arXiv:2409.18944} \BibitemShut
  {NoStop}%
\bibitem [{\citenamefont {Wang}\ \emph {et~al.}(2026)\citenamefont {Wang},
  \citenamefont {Kiely}, \citenamefont {Tell}, \citenamefont {Obermeyer},
  \citenamefont {Barendregt}, \citenamefont {Bojović}, \citenamefont {Preiss},
  \citenamefont {Sarma}, \citenamefont {Franz}, \citenamefont {Fisher},
  \citenamefont {Xu},\ and\ \citenamefont {Bloch}}]{IBloch2026}%
  \BibitemOpen
  \bibfield  {author} {\bibinfo {author} {\bibfnamefont {S.}~\bibnamefont
  {Wang}}, \bibinfo {author} {\bibfnamefont {T.~G.}\ \bibnamefont {Kiely}},
  \bibinfo {author} {\bibfnamefont {D.}~\bibnamefont {Tell}}, \bibinfo {author}
  {\bibfnamefont {J.}~\bibnamefont {Obermeyer}}, \bibinfo {author}
  {\bibfnamefont {M.}~\bibnamefont {Barendregt}}, \bibinfo {author}
  {\bibfnamefont {P.}~\bibnamefont {Bojović}}, \bibinfo {author}
  {\bibfnamefont {P.~M.}\ \bibnamefont {Preiss}}, \bibinfo {author}
  {\bibfnamefont {A.}~\bibnamefont {Sarma}}, \bibinfo {author} {\bibfnamefont
  {T.}~\bibnamefont {Franz}}, \bibinfo {author} {\bibfnamefont {M.~P.~A.}\
  \bibnamefont {Fisher}}, \bibinfo {author} {\bibfnamefont {C.}~\bibnamefont
  {Xu}},\ and\ \bibinfo {author} {\bibfnamefont {I.}~\bibnamefont {Bloch}},\
  }\href@noop {} {\bibinfo {title} {Observation of strong-to-weak spontaneous
  symmetry breaking in a dephased fermi gas}} (\bibinfo {year} {2026}),\
  \Eprint {https://arxiv.org/abs/arXiv:2604.16137} {arXiv:2604.16137}
  \BibitemShut {NoStop}%
\bibitem [{\citenamefont {Lu}\ \emph {et~al.}(2026)\citenamefont {Lu},
  \citenamefont {Liu}, \citenamefont {Gopalakrishnan},\ and\ \citenamefont
  {You}}]{YZYou2026PRXQ}%
  \BibitemOpen
  \bibfield  {author} {\bibinfo {author} {\bibfnamefont {T.-C.}\ \bibnamefont
  {Lu}}, \bibinfo {author} {\bibfnamefont {Y.-J.}\ \bibnamefont {Liu}},
  \bibinfo {author} {\bibfnamefont {S.}~\bibnamefont {Gopalakrishnan}},\ and\
  \bibinfo {author} {\bibfnamefont {Y.}~\bibnamefont {You}},\ }\bibfield
  {title} {\bibinfo {title} {Holographic duality between bulk topological order
  and boundary mixed-state order},\ }\href {https://doi.org/10.1103/lrqp-pnl1}
  {\bibfield  {journal} {\bibinfo  {journal} {PRX Quantum}\ }\textbf {\bibinfo
  {volume} {7}},\ \bibinfo {pages} {033041} (\bibinfo {year}
  {2026})}\BibitemShut {NoStop}%
\bibitem [{\citenamefont {Hauser}\ \emph {et~al.}(2026)\citenamefont {Hauser},
  \citenamefont {Su}, \citenamefont {Ha}, \citenamefont {Lloyd}, \citenamefont
  {Kiely}, \citenamefont {Vasseur}, \citenamefont {Gopalakrishnan},
  \citenamefont {Xu},\ and\ \citenamefont {Fisher}}]{MFisher2026arXiv}%
  \BibitemOpen
  \bibfield  {author} {\bibinfo {author} {\bibfnamefont {J.}~\bibnamefont
  {Hauser}}, \bibinfo {author} {\bibfnamefont {K.}~\bibnamefont {Su}}, \bibinfo
  {author} {\bibfnamefont {H.}~\bibnamefont {Ha}}, \bibinfo {author}
  {\bibfnamefont {J.}~\bibnamefont {Lloyd}}, \bibinfo {author} {\bibfnamefont
  {T.~G.}\ \bibnamefont {Kiely}}, \bibinfo {author} {\bibfnamefont
  {R.}~\bibnamefont {Vasseur}}, \bibinfo {author} {\bibfnamefont
  {S.}~\bibnamefont {Gopalakrishnan}}, \bibinfo {author} {\bibfnamefont
  {C.}~\bibnamefont {Xu}},\ and\ \bibinfo {author} {\bibfnamefont {M.~P.~A.}\
  \bibnamefont {Fisher}},\ }\href@noop {} {\bibinfo {title} {Strong-to-weak
  symmetry breaking in open quantum systems: From discrete particles to
  continuum hydrodynamics}} (\bibinfo {year} {2026}),\ \Eprint
  {https://arxiv.org/abs/arXiv:2602.16045} {arXiv:2602.16045} \BibitemShut
  {NoStop}%
\bibitem [{\citenamefont {Shu}\ \emph {et~al.}(2026)\citenamefont {Shu},
  \citenamefont {Zhang}, \citenamefont {Luo}, \citenamefont {You},\ and\
  \citenamefont {Sun}}]{Chang2026arXiv}%
  \BibitemOpen
  \bibfield  {author} {\bibinfo {author} {\bibfnamefont {C.}~\bibnamefont
  {Shu}}, \bibinfo {author} {\bibfnamefont {K.}~\bibnamefont {Zhang}}, \bibinfo
  {author} {\bibfnamefont {Z.-X.}\ \bibnamefont {Luo}}, \bibinfo {author}
  {\bibfnamefont {Y.}~\bibnamefont {You}},\ and\ \bibinfo {author}
  {\bibfnamefont {K.}~\bibnamefont {Sun}},\ }\href@noop {} {\bibinfo {title}
  {Universal dynamical scaling of strong-to-weak spontaneous symmetry breaking
  in open quantum systems}} (\bibinfo {year} {2026}),\ \Eprint
  {https://arxiv.org/abs/arXiv:2603.06363} {arXiv:2603.06363} \BibitemShut
  {NoStop}%
\bibitem [{\citenamefont {Zhang}(2026)}]{Carolyn2026arXiv}%
  \BibitemOpen
  \bibfield  {author} {\bibinfo {author} {\bibfnamefont {C.}~\bibnamefont
  {Zhang}},\ }\href {https://doi.org/10.48550/arXiv.2605.29113} {\bibinfo
  {title} {Local diagnostics for strong-to-weak spontaneous symmetry breaking
  and non-equilibrium phase transitions}} (\bibinfo {year} {2026}),\ \Eprint
  {https://arxiv.org/abs/2605.29113} {arXiv:2605.29113 [quant-ph]} \BibitemShut
  {NoStop}%
\bibitem [{\citenamefont {Haga}\ \emph {et~al.}(2026)\citenamefont {Haga},
  \citenamefont {Kunimi},\ and\ \citenamefont {Nakagawa}}]{Haga2026arXiv}%
  \BibitemOpen
  \bibfield  {author} {\bibinfo {author} {\bibfnamefont {T.}~\bibnamefont
  {Haga}}, \bibinfo {author} {\bibfnamefont {M.}~\bibnamefont {Kunimi}},\ and\
  \bibinfo {author} {\bibfnamefont {M.}~\bibnamefont {Nakagawa}},\ }\href
  {https://doi.org/10.48550/arXiv.2606.15062} {\bibinfo {title} {Discontinuous
  strong-to-weak symmetry breaking transition from thermal pure states}}
  (\bibinfo {year} {2026}),\ \Eprint {https://arxiv.org/abs/2606.15062}
  {arXiv:2606.15062 [quant-ph]} \BibitemShut {NoStop}%
\bibitem [{\citenamefont {Chen}\ \emph {et~al.}(2025)\citenamefont {Chen},
  \citenamefont {Sun},\ and\ \citenamefont {Zhang}}]{PFZhang2025PRB}%
  \BibitemOpen
  \bibfield  {author} {\bibinfo {author} {\bibfnamefont {L.}~\bibnamefont
  {Chen}}, \bibinfo {author} {\bibfnamefont {N.}~\bibnamefont {Sun}},\ and\
  \bibinfo {author} {\bibfnamefont {P.}~\bibnamefont {Zhang}},\ }\bibfield
  {title} {\bibinfo {title} {Strong-to-weak symmetry breaking and entanglement
  transitions},\ }\href {https://doi.org/10.1103/PhysRevB.111.L060304}
  {\bibfield  {journal} {\bibinfo  {journal} {Phys. Rev. B}\ }\textbf {\bibinfo
  {volume} {111}},\ \bibinfo {pages} {L060304} (\bibinfo {year}
  {2025})}\BibitemShut {NoStop}%
\bibitem [{\citenamefont {Lu}\ \emph {et~al.}(2023)\citenamefont {Lu},
  \citenamefont {Zhang}, \citenamefont {Vijay},\ and\ \citenamefont
  {Hsieh}}]{HsiehT2023PRXQ}%
  \BibitemOpen
  \bibfield  {author} {\bibinfo {author} {\bibfnamefont {T.-C.}\ \bibnamefont
  {Lu}}, \bibinfo {author} {\bibfnamefont {Z.}~\bibnamefont {Zhang}}, \bibinfo
  {author} {\bibfnamefont {S.}~\bibnamefont {Vijay}},\ and\ \bibinfo {author}
  {\bibfnamefont {T.~H.}\ \bibnamefont {Hsieh}},\ }\bibfield  {title} {\bibinfo
  {title} {Mixed-state long-range order and criticality from measurement and
  feedback},\ }\href {https://doi.org/10.1103/PRXQuantum.4.030318} {\bibfield
  {journal} {\bibinfo  {journal} {PRX Quantum}\ }\textbf {\bibinfo {volume}
  {4}},\ \bibinfo {pages} {030318} (\bibinfo {year} {2023})}\BibitemShut
  {NoStop}%
\bibitem [{\citenamefont {Sang}\ \emph {et~al.}(2024)\citenamefont {Sang},
  \citenamefont {Zou},\ and\ \citenamefont {Hsieh}}]{Sang2024PRX}%
  \BibitemOpen
  \bibfield  {author} {\bibinfo {author} {\bibfnamefont {S.}~\bibnamefont
  {Sang}}, \bibinfo {author} {\bibfnamefont {Y.}~\bibnamefont {Zou}},\ and\
  \bibinfo {author} {\bibfnamefont {T.~H.}\ \bibnamefont {Hsieh}},\ }\bibfield
  {title} {\bibinfo {title} {Mixed-state quantum phases: Renormalization and
  quantum error correction},\ }\href
  {https://doi.org/10.1103/PhysRevX.14.031044} {\bibfield  {journal} {\bibinfo
  {journal} {Phys. Rev. X}\ }\textbf {\bibinfo {volume} {14}},\ \bibinfo
  {pages} {031044} (\bibinfo {year} {2024})}\BibitemShut {NoStop}%
\bibitem [{\citenamefont {Sang}\ and\ \citenamefont
  {Hsieh}(2025)}]{Hsieh2025PRL}%
  \BibitemOpen
  \bibfield  {author} {\bibinfo {author} {\bibfnamefont {S.}~\bibnamefont
  {Sang}}\ and\ \bibinfo {author} {\bibfnamefont {T.~H.}\ \bibnamefont
  {Hsieh}},\ }\bibfield  {title} {\bibinfo {title} {Stability of mixed-state
  quantum phases via finite markov length},\ }\href
  {https://doi.org/10.1103/PhysRevLett.134.070403} {\bibfield  {journal}
  {\bibinfo  {journal} {Phys. Rev. Lett.}\ }\textbf {\bibinfo {volume} {134}},\
  \bibinfo {pages} {070403} (\bibinfo {year} {2025})}\BibitemShut {NoStop}%
\bibitem [{\citenamefont {Zhu}\ \emph {et~al.}(2023)\citenamefont {Zhu},
  \citenamefont {Tantivasadakarn}, \citenamefont {Vishwanath}, \citenamefont
  {Trebst},\ and\ \citenamefont {Verresen}}]{Verresen2023PRL}%
  \BibitemOpen
  \bibfield  {author} {\bibinfo {author} {\bibfnamefont {G.-Y.}\ \bibnamefont
  {Zhu}}, \bibinfo {author} {\bibfnamefont {N.}~\bibnamefont
  {Tantivasadakarn}}, \bibinfo {author} {\bibfnamefont {A.}~\bibnamefont
  {Vishwanath}}, \bibinfo {author} {\bibfnamefont {S.}~\bibnamefont {Trebst}},\
  and\ \bibinfo {author} {\bibfnamefont {R.}~\bibnamefont {Verresen}},\
  }\bibfield  {title} {\bibinfo {title} {Nishimori's cat: Stable long-range
  entanglement from finite-depth unitaries and weak measurements},\ }\href
  {https://doi.org/10.1103/PhysRevLett.131.200201} {\bibfield  {journal}
  {\bibinfo  {journal} {Phys. Rev. Lett.}\ }\textbf {\bibinfo {volume} {131}},\
  \bibinfo {pages} {200201} (\bibinfo {year} {2023})}\BibitemShut {NoStop}%
\bibitem [{\citenamefont {Li}\ and\ \citenamefont
  {Fisher}(2023)}]{MatthewF2023PRB}%
  \BibitemOpen
  \bibfield  {author} {\bibinfo {author} {\bibfnamefont {Y.}~\bibnamefont
  {Li}}\ and\ \bibinfo {author} {\bibfnamefont {M.~P.~A.}\ \bibnamefont
  {Fisher}},\ }\bibfield  {title} {\bibinfo {title} {Decodable hybrid dynamics
  of open quantum systems with ${\mathbb{z}}_{2}$ symmetry},\ }\href
  {https://doi.org/10.1103/PhysRevB.108.214302} {\bibfield  {journal} {\bibinfo
   {journal} {Phys. Rev. B}\ }\textbf {\bibinfo {volume} {108}},\ \bibinfo
  {pages} {214302} (\bibinfo {year} {2023})}\BibitemShut {NoStop}%
\bibitem [{\citenamefont {Lee}\ \emph {et~al.}(2024)\citenamefont {Lee},
  \citenamefont {Oh}, \citenamefont {Wong}, \citenamefont {Chen},\ and\
  \citenamefont {Jiang}}]{JiangL2024PRL}%
  \BibitemOpen
  \bibfield  {author} {\bibinfo {author} {\bibfnamefont {S.-u.}\ \bibnamefont
  {Lee}}, \bibinfo {author} {\bibfnamefont {C.}~\bibnamefont {Oh}}, \bibinfo
  {author} {\bibfnamefont {Y.}~\bibnamefont {Wong}}, \bibinfo {author}
  {\bibfnamefont {S.}~\bibnamefont {Chen}},\ and\ \bibinfo {author}
  {\bibfnamefont {L.}~\bibnamefont {Jiang}},\ }\bibfield  {title} {\bibinfo
  {title} {Universal spreading of conditional mutual information in noisy
  random circuits},\ }\href {https://doi.org/10.1103/PhysRevLett.133.200402}
  {\bibfield  {journal} {\bibinfo  {journal} {Phys. Rev. Lett.}\ }\textbf
  {\bibinfo {volume} {133}},\ \bibinfo {pages} {200402} (\bibinfo {year}
  {2024})}\BibitemShut {NoStop}%
\bibitem [{\citenamefont {Zhang}\ and\ \citenamefont
  {Gopalakrishnan}(2024)}]{ZhangYF2024PRA}%
  \BibitemOpen
  \bibfield  {author} {\bibinfo {author} {\bibfnamefont {Y.}~\bibnamefont
  {Zhang}}\ and\ \bibinfo {author} {\bibfnamefont {S.}~\bibnamefont
  {Gopalakrishnan}},\ }\bibfield  {title} {\bibinfo {title} {Nonlocal growth of
  quantum conditional mutual information under decoherence},\ }\href
  {https://doi.org/10.1103/PhysRevA.110.032426} {\bibfield  {journal} {\bibinfo
   {journal} {Phys. Rev. A}\ }\textbf {\bibinfo {volume} {110}},\ \bibinfo
  {pages} {032426} (\bibinfo {year} {2024})}\BibitemShut {NoStop}%
\bibitem [{\citenamefont {Zhang}\ and\ \citenamefont
  {Gopalakrishnan}(2025)}]{ZhangYF2025PRL}%
  \BibitemOpen
  \bibfield  {author} {\bibinfo {author} {\bibfnamefont {Y.~F.}\ \bibnamefont
  {Zhang}}\ and\ \bibinfo {author} {\bibfnamefont {S.}~\bibnamefont
  {Gopalakrishnan}},\ }\bibfield  {title} {\bibinfo {title} {Conditional mutual
  information and information-theoretic phases of decohered gibbs states},\
  }\href {https://doi.org/10.1103/cjkh-1z6c} {\bibfield  {journal} {\bibinfo
  {journal} {Phys. Rev. Lett.}\ }\textbf {\bibinfo {volume} {135}},\ \bibinfo
  {pages} {160401} (\bibinfo {year} {2025})}\BibitemShut {NoStop}%
\bibitem [{\citenamefont {Lee}(2025)}]{LeeJY2025PRL}%
  \BibitemOpen
  \bibfield  {author} {\bibinfo {author} {\bibfnamefont {J.~Y.}\ \bibnamefont
  {Lee}},\ }\bibfield  {title} {\bibinfo {title} {Exact calculations of
  coherent information for toric codes under decoherence: Identifying the
  fundamental error threshold},\ }\href {https://doi.org/10.1103/hlfh-86yz}
  {\bibfield  {journal} {\bibinfo  {journal} {Phys. Rev. Lett.}\ }\textbf
  {\bibinfo {volume} {134}},\ \bibinfo {pages} {250601} (\bibinfo {year}
  {2025})}\BibitemShut {NoStop}%
\bibitem [{\citenamefont {Sang}\ \emph {et~al.}(2025)\citenamefont {Sang},
  \citenamefont {Lessa}, \citenamefont {Mong}, \citenamefont {Grover},
  \citenamefont {Wang},\ and\ \citenamefont {Hsieh}}]{Timothy2025}%
  \BibitemOpen
  \bibfield  {author} {\bibinfo {author} {\bibfnamefont {S.}~\bibnamefont
  {Sang}}, \bibinfo {author} {\bibfnamefont {L.~A.}\ \bibnamefont {Lessa}},
  \bibinfo {author} {\bibfnamefont {R.~S.~K.}\ \bibnamefont {Mong}}, \bibinfo
  {author} {\bibfnamefont {T.}~\bibnamefont {Grover}}, \bibinfo {author}
  {\bibfnamefont {C.}~\bibnamefont {Wang}},\ and\ \bibinfo {author}
  {\bibfnamefont {T.~H.}\ \bibnamefont {Hsieh}},\ }\href@noop {} {\bibinfo
  {title} {Mixed-state phases from local reversibility}} (\bibinfo {year}
  {2025}),\ \Eprint {https://arxiv.org/abs/arXiv:2507.02292} {arXiv:2507.02292}
  \BibitemShut {NoStop}%
\bibitem [{\citenamefont {Lee}(2026)}]{JongYLee2026arXiv}%
  \BibitemOpen
  \bibfield  {author} {\bibinfo {author} {\bibfnamefont {J.~Y.}\ \bibnamefont
  {Lee}},\ }\href@noop {} {\bibinfo {title} {Charge scrambling in
  strong-to-weak spontaneous symmetry breaking}} (\bibinfo {year} {2026}),\
  \Eprint {https://arxiv.org/abs/arXiv:2605.05288} {arXiv:2605.05288}
  \BibitemShut {NoStop}%
\bibitem [{\citenamefont {Kumar}\ \emph {et~al.}(2026)\citenamefont {Kumar},
  \citenamefont {Zou}, \citenamefont {Negari}, \citenamefont {Melko},\ and\
  \citenamefont {Hsieh}}]{TimothyHH2026arXiv}%
  \BibitemOpen
  \bibfield  {author} {\bibinfo {author} {\bibfnamefont {T.~A.}\ \bibnamefont
  {Kumar}}, \bibinfo {author} {\bibfnamefont {Y.}~\bibnamefont {Zou}}, \bibinfo
  {author} {\bibfnamefont {A.-R.}\ \bibnamefont {Negari}}, \bibinfo {author}
  {\bibfnamefont {R.~G.}\ \bibnamefont {Melko}},\ and\ \bibinfo {author}
  {\bibfnamefont {T.~H.}\ \bibnamefont {Hsieh}},\ }\href@noop {} {\bibinfo
  {title} {Unlearnable phases of matter}} (\bibinfo {year} {2026}),\ \Eprint
  {https://arxiv.org/abs/arXiv:2602.11262} {arXiv:2602.11262} \BibitemShut
  {NoStop}%
\bibitem [{\citenamefont {Divi}\ \emph {et~al.}(2026)\citenamefont {Divi},
  \citenamefont {Lessa},\ and\ \citenamefont {Wang}}]{Wang2026LocalSWSSB}%
  \BibitemOpen
  \bibfield  {author} {\bibinfo {author} {\bibfnamefont {F.}~\bibnamefont
  {Divi}}, \bibinfo {author} {\bibfnamefont {L.~A.}\ \bibnamefont {Lessa}},\
  and\ \bibinfo {author} {\bibfnamefont {C.}~\bibnamefont {Wang}},\ }\href@noop
  {} {\bibinfo {title} {Local strong-to-weak spontaneous symmetry breaking}}
  (\bibinfo {year} {2026}),\ \Eprint {https://arxiv.org/abs/2605.28967}
  {arXiv:2605.28967 [quant-ph]} \BibitemShut {NoStop}%
\bibitem [{\citenamefont {Liu}\ \emph {et~al.}(2026)\citenamefont {Liu},
  \citenamefont {Yi},\ and\ \citenamefont {Else}}]{LiuRZ2026arXiv}%
  \BibitemOpen
  \bibfield  {author} {\bibinfo {author} {\bibfnamefont {R.}~\bibnamefont
  {Liu}}, \bibinfo {author} {\bibfnamefont {J.}~\bibnamefont {Yi}},\ and\
  \bibinfo {author} {\bibfnamefont {D.~V.}\ \bibnamefont {Else}},\ }\href@noop
  {} {\bibinfo {title} {A local description of strong symmetries and
  strong-to-weak symmetry breaking in quantum many-body systems}} (\bibinfo
  {year} {2026}),\ \Eprint {https://arxiv.org/abs/arXiv:2605.28925}
  {arXiv:2605.28925} \BibitemShut {NoStop}%
\bibitem [{\citenamefont {Zhao}\ \emph {et~al.}(2026)\citenamefont {Zhao},
  \citenamefont {Rao},\ and\ \citenamefont {Liu}}]{DongELiu2026PRB}%
  \BibitemOpen
  \bibfield  {author} {\bibinfo {author} {\bibfnamefont {Y.}~\bibnamefont
  {Zhao}}, \bibinfo {author} {\bibfnamefont {L.}~\bibnamefont {Rao}},\ and\
  \bibinfo {author} {\bibfnamefont {D.~E.}\ \bibnamefont {Liu}},\ }\bibfield
  {title} {\bibinfo {title} {Noncommutative weak measurements: Entanglement,
  symmetry breaking, and the role of readout},\ }\href
  {https://doi.org/10.1103/q7hz-qcxz} {\bibfield  {journal} {\bibinfo
  {journal} {Phys. Rev. B}\ }\textbf {\bibinfo {volume} {113}},\ \bibinfo
  {pages} {174312} (\bibinfo {year} {2026})}\BibitemShut {NoStop}%
\bibitem [{\citenamefont {Fawzi}\ and\ \citenamefont
  {Renner}(2015)}]{Omar2015CMP}%
  \BibitemOpen
  \bibfield  {author} {\bibinfo {author} {\bibfnamefont {O.}~\bibnamefont
  {Fawzi}}\ and\ \bibinfo {author} {\bibfnamefont {R.}~\bibnamefont {Renner}},\
  }\bibfield  {title} {\bibinfo {title} {Quantum conditional mutual information
  and approximate markov chains},\ }\href
  {https://doi.org/10.1007/s00220-015-2466-x} {\bibfield  {journal} {\bibinfo
  {journal} {Communications in Mathematical Physics}\ }\textbf {\bibinfo
  {volume} {340}},\ \bibinfo {pages} {575} (\bibinfo {year}
  {2015})}\BibitemShut {NoStop}%
\bibitem [{\citenamefont {Agrawal}\ \emph {et~al.}(2022)\citenamefont
  {Agrawal}, \citenamefont {Zabalo}, \citenamefont {Chen}, \citenamefont
  {Wilson}, \citenamefont {Potter}, \citenamefont {Pixley}, \citenamefont
  {Gopalakrishnan},\ and\ \citenamefont {Vasseur}}]{Vasseur2022PRX}%
  \BibitemOpen
  \bibfield  {author} {\bibinfo {author} {\bibfnamefont {U.}~\bibnamefont
  {Agrawal}}, \bibinfo {author} {\bibfnamefont {A.}~\bibnamefont {Zabalo}},
  \bibinfo {author} {\bibfnamefont {K.}~\bibnamefont {Chen}}, \bibinfo {author}
  {\bibfnamefont {J.~H.}\ \bibnamefont {Wilson}}, \bibinfo {author}
  {\bibfnamefont {A.~C.}\ \bibnamefont {Potter}}, \bibinfo {author}
  {\bibfnamefont {J.~H.}\ \bibnamefont {Pixley}}, \bibinfo {author}
  {\bibfnamefont {S.}~\bibnamefont {Gopalakrishnan}},\ and\ \bibinfo {author}
  {\bibfnamefont {R.}~\bibnamefont {Vasseur}},\ }\bibfield  {title} {\bibinfo
  {title} {Entanglement and charge-sharpening transitions in u(1) symmetric
  monitored quantum circuits},\ }\href
  {https://doi.org/10.1103/PhysRevX.12.041002} {\bibfield  {journal} {\bibinfo
  {journal} {Phys. Rev. X}\ }\textbf {\bibinfo {volume} {12}},\ \bibinfo
  {pages} {041002} (\bibinfo {year} {2022})}\BibitemShut {NoStop}%
\bibitem [{\citenamefont {Zerba}\ \emph {et~al.}(2025)\citenamefont {Zerba},
  \citenamefont {Gopalakrishnan},\ and\ \citenamefont {Knap}}]{Knap2025arXiv}%
  \BibitemOpen
  \bibfield  {author} {\bibinfo {author} {\bibfnamefont {C.}~\bibnamefont
  {Zerba}}, \bibinfo {author} {\bibfnamefont {S.}~\bibnamefont
  {Gopalakrishnan}},\ and\ \bibinfo {author} {\bibfnamefont {M.}~\bibnamefont
  {Knap}},\ }\href@noop {} {\bibinfo {title} {Strong-to-weak symmetry breaking
  in monitored dipole conserving quantum circuits}} (\bibinfo {year} {2025}),\
  \Eprint {https://arxiv.org/abs/arXiv:2512.14830} {arXiv:2512.14830}
  \BibitemShut {NoStop}%
\bibitem [{\citenamefont {Mallick}(2015)}]{Mallick2015}%
  \BibitemOpen
  \bibfield  {author} {\bibinfo {author} {\bibfnamefont {K.}~\bibnamefont
  {Mallick}},\ }\bibfield  {title} {\bibinfo {title} {The exclusion process: A
  paradigm for non-equilibrium behaviour},\ }\href
  {https://doi.org/https://doi.org/10.1016/j.physa.2014.07.046} {\bibfield
  {journal} {\bibinfo  {journal} {Physica A: Statistical Mechanics and its
  Applications}\ }\textbf {\bibinfo {volume} {418}},\ \bibinfo {pages} {17}
  (\bibinfo {year} {2015})},\ \bibinfo {note} {proceedings of the 13th
  International Summer School on Fundamental Problems in Statistical
  Physics}\BibitemShut {NoStop}%
\bibitem [{\citenamefont {Bernard}\ \emph {et~al.}(2018)\citenamefont
  {Bernard}, \citenamefont {Jin},\ and\ \citenamefont
  {Shpielberg}}]{Bernard2018}%
  \BibitemOpen
  \bibfield  {author} {\bibinfo {author} {\bibfnamefont {D.}~\bibnamefont
  {Bernard}}, \bibinfo {author} {\bibfnamefont {T.}~\bibnamefont {Jin}},\ and\
  \bibinfo {author} {\bibfnamefont {O.}~\bibnamefont {Shpielberg}},\ }\bibfield
   {title} {\bibinfo {title} {Transport in quantum chains under strong
  monitoring},\ }\href {https://doi.org/10.1209/0295-5075/121/60006} {\bibfield
   {journal} {\bibinfo  {journal} {Europhysics Letters}\ }\textbf {\bibinfo
  {volume} {121}},\ \bibinfo {pages} {60006} (\bibinfo {year}
  {2018})}\BibitemShut {NoStop}%
\bibitem [{\citenamefont {Wellnitz}\ \emph {et~al.}(2022)\citenamefont
  {Wellnitz}, \citenamefont {Preisser}, \citenamefont {Alba}, \citenamefont
  {Dubail},\ and\ \citenamefont {Schachenmayer}}]{Wellnitz2022PRL}%
  \BibitemOpen
  \bibfield  {author} {\bibinfo {author} {\bibfnamefont {D.}~\bibnamefont
  {Wellnitz}}, \bibinfo {author} {\bibfnamefont {G.}~\bibnamefont {Preisser}},
  \bibinfo {author} {\bibfnamefont {V.}~\bibnamefont {Alba}}, \bibinfo {author}
  {\bibfnamefont {J.}~\bibnamefont {Dubail}},\ and\ \bibinfo {author}
  {\bibfnamefont {J.}~\bibnamefont {Schachenmayer}},\ }\bibfield  {title}
  {\bibinfo {title} {Rise and fall, and slow rise again, of operator
  entanglement under dephasing},\ }\href
  {https://doi.org/10.1103/PhysRevLett.129.170401} {\bibfield  {journal}
  {\bibinfo  {journal} {Phys. Rev. Lett.}\ }\textbf {\bibinfo {volume} {129}},\
  \bibinfo {pages} {170401} (\bibinfo {year} {2022})}\BibitemShut {NoStop}%
\bibitem [{\citenamefont {Feldmeier}\ \emph {et~al.}(2020)\citenamefont
  {Feldmeier}, \citenamefont {Sala}, \citenamefont {De~Tomasi}, \citenamefont
  {Pollmann},\ and\ \citenamefont {Knap}}]{Knap2020PRL}%
  \BibitemOpen
  \bibfield  {author} {\bibinfo {author} {\bibfnamefont {J.}~\bibnamefont
  {Feldmeier}}, \bibinfo {author} {\bibfnamefont {P.}~\bibnamefont {Sala}},
  \bibinfo {author} {\bibfnamefont {G.}~\bibnamefont {De~Tomasi}}, \bibinfo
  {author} {\bibfnamefont {F.}~\bibnamefont {Pollmann}},\ and\ \bibinfo
  {author} {\bibfnamefont {M.}~\bibnamefont {Knap}},\ }\bibfield  {title}
  {\bibinfo {title} {Anomalous diffusion in dipole- and
  higher-moment-conserving systems},\ }\href
  {https://doi.org/10.1103/PhysRevLett.125.245303} {\bibfield  {journal}
  {\bibinfo  {journal} {Phys. Rev. Lett.}\ }\textbf {\bibinfo {volume} {125}},\
  \bibinfo {pages} {245303} (\bibinfo {year} {2020})}\BibitemShut {NoStop}%
\bibitem [{\citenamefont {Ogunnaike}\ \emph {et~al.}(2023)\citenamefont
  {Ogunnaike}, \citenamefont {Feldmeier},\ and\ \citenamefont
  {Lee}}]{JoneYeon2023PRL}%
  \BibitemOpen
  \bibfield  {author} {\bibinfo {author} {\bibfnamefont {O.}~\bibnamefont
  {Ogunnaike}}, \bibinfo {author} {\bibfnamefont {J.}~\bibnamefont
  {Feldmeier}},\ and\ \bibinfo {author} {\bibfnamefont {J.~Y.}\ \bibnamefont
  {Lee}},\ }\bibfield  {title} {\bibinfo {title} {Unifying emergent
  hydrodynamics and lindbladian low-energy spectra across symmetries,
  constraints, and long-range interactions},\ }\href
  {https://doi.org/10.1103/PhysRevLett.131.220403} {\bibfield  {journal}
  {\bibinfo  {journal} {Phys. Rev. Lett.}\ }\textbf {\bibinfo {volume} {131}},\
  \bibinfo {pages} {220403} (\bibinfo {year} {2023})}\BibitemShut {NoStop}%
\bibitem [{\citenamefont {Moudgalya}\ and\ \citenamefont
  {Motrunich}(2024)}]{Moudgalya2024PRXQ}%
  \BibitemOpen
  \bibfield  {author} {\bibinfo {author} {\bibfnamefont {S.}~\bibnamefont
  {Moudgalya}}\ and\ \bibinfo {author} {\bibfnamefont {O.~I.}\ \bibnamefont
  {Motrunich}},\ }\bibfield  {title} {\bibinfo {title} {Symmetries as ground
  states of local superoperators: Hydrodynamic implications},\ }\href
  {https://doi.org/10.1103/PRXQuantum.5.040330} {\bibfield  {journal} {\bibinfo
   {journal} {PRX Quantum}\ }\textbf {\bibinfo {volume} {5}},\ \bibinfo {pages}
  {040330} (\bibinfo {year} {2024})}\BibitemShut {NoStop}%
\bibitem [{\citenamefont {Huang}\ \emph
  {et~al.}(2025{\natexlab{b}})\citenamefont {Huang}, \citenamefont {Qi},
  \citenamefont {Zhang},\ and\ \citenamefont {Lucas}}]{Lucas2025PRB}%
  \BibitemOpen
  \bibfield  {author} {\bibinfo {author} {\bibfnamefont {X.}~\bibnamefont
  {Huang}}, \bibinfo {author} {\bibfnamefont {M.}~\bibnamefont {Qi}}, \bibinfo
  {author} {\bibfnamefont {J.-H.}\ \bibnamefont {Zhang}},\ and\ \bibinfo
  {author} {\bibfnamefont {A.}~\bibnamefont {Lucas}},\ }\bibfield  {title}
  {\bibinfo {title} {Hydrodynamics as the effective field theory of
  strong-to-weak spontaneous symmetry breaking},\ }\href
  {https://doi.org/10.1103/PhysRevB.111.125147} {\bibfield  {journal} {\bibinfo
   {journal} {Phys. Rev. B}\ }\textbf {\bibinfo {volume} {111}},\ \bibinfo
  {pages} {125147} (\bibinfo {year} {2025}{\natexlab{b}})}\BibitemShut
  {NoStop}%
\bibitem [{Sup()}]{SupMat}%
  \BibitemOpen
  \href@noop {} {}\bibinfo {note} {See Supplemental Material for more
  details}\BibitemShut {NoStop}%
\bibitem [{foo()}]{footnote1}%
  \BibitemOpen
  \href@noop {} {}\bibinfo {note} {For a strongly symmetric state $\rho$ and a
  charged local operator $O_i$, we have $P_Q \rho^{k/2} P_Q=\rho^{k/2}$ and
  $P_Q O_i P_Q=0$, where $P_Q$ denotes the projector onto the charge-$Q$
  sector. Hence, $\Tr[\rho^{k/2}O_i\rho^{k/2}O_i^\dagger]=\Tr[P_Q\rho^{k/2}P_Q
  O_i P_Q \rho^{k/2} P_Q O_i^\dagger] = 0$}\BibitemShut {NoStop}%
\end{thebibliography}

\begin{thebibliography}{3}%
\makeatletter
\providecommand \@ifxundefined [1]{%
 \@ifx{#1\undefined}
}%
\providecommand \@ifnum [1]{%
 \ifnum #1\expandafter \@firstoftwo
 \else \expandafter \@secondoftwo
 \fi
}%
\providecommand \@ifx [1]{%
 \ifx #1\expandafter \@firstoftwo
 \else \expandafter \@secondoftwo
 \fi
}%
\providecommand \natexlab [1]{#1}%
\providecommand \enquote  [1]{``#1''}%
\providecommand \bibnamefont  [1]{#1}%
\providecommand \bibfnamefont [1]{#1}%
\providecommand \citenamefont [1]{#1}%
\providecommand \href@noop [0]{\@secondoftwo}%
\providecommand \href [0]{\begingroup \@sanitize@url \@href}%
\providecommand \@href[1]{\@@startlink{#1}\@@href}%
\providecommand \@@href[1]{\endgroup#1\@@endlink}%
\providecommand \@sanitize@url [0]{\catcode `\\12\catcode `\$12\catcode `\&12\catcode `\#12\catcode `\^12\catcode `\_12\catcode `\%12\relax}%
\providecommand \@@startlink[1]{}%
\providecommand \@@endlink[0]{}%
\providecommand \url  [0]{\begingroup\@sanitize@url \@url }%
\providecommand \@url [1]{\endgroup\@href {#1}{\urlprefix }}%
\providecommand \urlprefix  [0]{URL }%
\providecommand \Eprint [0]{\href }%
\providecommand \doibase [0]{https://doi.org/}%
\providecommand \selectlanguage [0]{\@gobble}%
\providecommand \bibinfo  [0]{\@secondoftwo}%
\providecommand \bibfield  [0]{\@secondoftwo}%
\providecommand \translation [1]{[#1]}%
\providecommand \BibitemOpen [0]{}%
\providecommand \bibitemStop [0]{}%
\providecommand \bibitemNoStop [0]{.\EOS\space}%
\providecommand \EOS [0]{\spacefactor3000\relax}%
\providecommand \BibitemShut  [1]{\csname bibitem#1\endcsname}%
\let\auto@bib@innerbib\@empty
\bibitem [{\citenamefont {Gromov}\ \emph {et~al.}(2020)\citenamefont {Gromov}, \citenamefont {Lucas},\ and\ \citenamefont {Nandkishore}}]{dipole_Hydrodynamics_PRR_2020}%
  \BibitemOpen
  \bibfield  {author} {\bibinfo {author} {\bibfnamefont {A.}~\bibnamefont {Gromov}}, \bibinfo {author} {\bibfnamefont {A.}~\bibnamefont {Lucas}},\ and\ \bibinfo {author} {\bibfnamefont {R.~M.}\ \bibnamefont {Nandkishore}},\ }\bibfield  {title} {\bibinfo {title} {Fracton hydrodynamics},\ }\href {https://doi.org/10.1103/PhysRevResearch.2.033124} {\bibfield  {journal} {\bibinfo  {journal} {Phys. Rev. Res.}\ }\textbf {\bibinfo {volume} {2}},\ \bibinfo {pages} {033124} (\bibinfo {year} {2020})}\BibitemShut {NoStop}%
\bibitem [{\citenamefont {Popkov}\ \emph {et~al.}(2015)\citenamefont {Popkov}, \citenamefont {Schadschneider}, \citenamefont {Schmidt},\ and\ \citenamefont {Schütz}}]{Dynamical_univer_class_PNAS_2015}%
  \BibitemOpen
  \bibfield  {author} {\bibinfo {author} {\bibfnamefont {V.}~\bibnamefont {Popkov}}, \bibinfo {author} {\bibfnamefont {A.}~\bibnamefont {Schadschneider}}, \bibinfo {author} {\bibfnamefont {J.}~\bibnamefont {Schmidt}},\ and\ \bibinfo {author} {\bibfnamefont {G.~M.}\ \bibnamefont {Schütz}},\ }\bibfield  {title} {\bibinfo {title} {Fibonacci family of dynamical universality classes},\ }\href {https://doi.org/10.1073/pnas.1512261112} {\bibfield  {journal} {\bibinfo  {journal} {Proceedings of the National Academy of Sciences}\ }\textbf {\bibinfo {volume} {112}},\ \bibinfo {pages} {12645} (\bibinfo {year} {2015})}\BibitemShut {NoStop}%
\bibitem [{\citenamefont {Gorenflo}\ \emph {et~al.}(2020)\citenamefont {Gorenflo}, \citenamefont {Kilbas}, \citenamefont {Mainardi}, \citenamefont {Rogosin} \emph {et~al.}}]{Gorenflo2020Book}%
  \BibitemOpen
  \bibfield  {author} {\bibinfo {author} {\bibfnamefont {R.}~\bibnamefont {Gorenflo}}, \bibinfo {author} {\bibfnamefont {A.~A.}\ \bibnamefont {Kilbas}}, \bibinfo {author} {\bibfnamefont {F.}~\bibnamefont {Mainardi}}, \bibinfo {author} {\bibfnamefont {S.~V.}\ \bibnamefont {Rogosin}}, \emph {et~al.},\ }\href@noop {} {\emph {\bibinfo {title} {Mittag-Leffler functions, related topics and applications}}},\ Vol.\ \bibinfo {volume} {1451}\ (\bibinfo  {publisher} {Springer},\ \bibinfo {year} {2020})\BibitemShut {NoStop}%
\end{thebibliography}
\end{document}